\documentclass[nofootinbib,showpacs]{revtex4}

\usepackage{graphicx}
\usepackage{dcolumn}
\usepackage{amsmath,amssymb,epsfig}
\usepackage{paralist}
\usepackage{comment}
\usepackage{graphicx}
\usepackage{multirow}
\usepackage{color}
\usepackage{soul}

\allowdisplaybreaks

\renewcommand{\vec}[1]{\boldsymbol{\mathrm{#1}}}

\begin{document}


\title{
Accelerating relativistic reference frames in Minkowski space-time}

\author{Slava G. Turyshev$^1$, Olivier L. Minazzoli$^1$, and Viktor T. Toth$^2$
}

\affiliation{\vskip 3pt
$^1$Jet Propulsion Laboratory, California Institute of Technology,\\
4800 Oak Grove Drive, Pasadena, CA 91109-0899, USA
}%

\affiliation{\vskip 3pt
$^2$Ottawa, ON, Canada,
}%

\date{\today}

\begin{abstract}

We study accelerating relativistic reference frames in Minkowski space-time under the harmonic gauge. It is well-known that the harmonic gauge imposes constraints on the components of the metric tensor and also on the functional form of admissible coordinate transformations. These two sets of constraints are equivalent and represent the dual nature of the harmonic gauge. We explore this duality and show that the harmonic gauge allows presenting an accelerated metric in an elegant form that depends only on two harmonic potentials. It also allows reconstruction of the spatial structure of the post-Galilean coordinate transformation functions relating inertial and accelerating frames. The remaining temporal dependence of these functions together with corresponding equations of motion are determined from dynamical conditions, obtained by constructing the relativistic proper reference frame of an accelerated test particle. In this frame, the effect of external forces acting on the observer is balanced by the fictitious frame-reaction force that is needed to keep the test particle at rest with respect to the frame, conserving its relativistic linear momentum. We find that this approach is sufficient to determine all the terms of the coordinate transformation. The same method is then used to develop the inverse transformations. The resulting post-Galilean coordinate transformations extend the Poincar\'e group on the case of accelerating observers. We present and discuss the resulting coordinate transformations, relativistic equations of motion, and the structure of the metric tensors corresponding to the relativistic reference frames involved.

\end{abstract}


\maketitle

\section{Introduction}

Most modern precision physics experiments are conducted in non-inertial reference frames corresponding to either the surface of the rotating Earth or an accelerating spacecraft \cite{Turyshev:2007qy,Turyshev:2008ur}. Quite often, in addition to special precautions aimed at reducing the ambient non-gravitational acceleration noise, one also needs to develop a relativistic treatment of all the observable quantities. Unfortunately, there is no general agreement on how to approach these problems in a relativistic manner. For instance, the  treatment advocated in \cite{MTW,Ni-1977,Ni-Zimmermann-1978} is incompatible with most of the methods of relativistic reference frames \cite{Brumberg-Kopejkin-1989,Kopejkin-1988,Brumberg-Kopejkin-1989-2,DSX-I,DSX-II,DSX-III,DSX-IV,Kopeikin:2004ia} and also is inadequate from the practical standpoint involving the transformation between the experimental (accelerated) and laboratory  (inertial) frames. In general, it is not clear how to write a suitable explicit form of the metric tensor of the accelerated reference frame and construct coordinate transformations between the inertial and accelerated frames, especially when high accuracy is required.

Within the realm of special theory of relativity, to describe the dynamics, one uses the relativistic mechanics of Poincar\'e which uses the Lorentz transformations to describe physical process in various inertial reference frames:
{}
\begin{eqnarray}
t'&=&\gamma\Big(t+{{\vec v}_0\cdot{\vec r}\over c^2}\Big),
\qquad
{\vec r}'={\vec r}+\gamma {\vec v}_0 t+(\gamma-1){{\vec v}_0({\vec v}_0\cdot{\vec r})\over v^2_0},
\qquad
\gamma =\Big(1-{v^2_0\over c^2}\Big)^{-{1\over2}}. \label{eq:1.7a}
\end{eqnarray}
Note that the transformations inverse  to those of (\ref{eq:1.7a}) are obtained by simply replacing $(t',{\vec r}')\rightarrow(t,{\vec r})$ and the velocity ${\vec v}_0\rightarrow-{\vec v}_0$. The Lorentz transformations are well suited to study the case of uniform motion with constant velocity. However, while analyzing experimental data one often has to deal with acceleration and dynamical noise whose presence limits the practical applicability of these results. To address the presence of acceleration, for each instant of time one can use (\ref{eq:1.7a}) to conceptually define a set of instantaneous quasi-inertial reference frames. Although useful conceptually, this approach is insufficient in practice as it leads to a decreased precision on larger time scales.

When the Riemannian geometry of general theory of relativity is concerned, it is well known that coordinate charts are merely labels. Usually, space-time coordinates have no direct physical meaning and it is essential to construct the observables as coordinate-independent quantities. Thus, in order to interpret the results of observations or experiments, one picks a specific coordinate system, chosen for the sake of convenience and calculational expediency, formulates a coordinate picture of the measurement procedure, and then one derives the observable out of it. It is also known that an ill-defined reference frame may lead to appearance of non-physical terms that may significantly complicate the interpretation of the data collected \cite{Brumberg-Kopejkin-1989}. Therefore, in practical problems involving relativistic reference frames, choosing the right coordinate system with clearly understood properties is of paramount importance, even as one recognizes that in principle, all (non-degenerate) coordinate systems are created equal \cite{Soffel:2003cr}.

Modern theories of relativistic reference frames \cite{Brumberg-Kopejkin-1989,Kopejkin-1988,Brumberg-Kopejkin-1989-2,DSX-I,DSX-II,DSX-III,DSX-IV,Kopeikin:2004ia}, dealing predominantly with the general theory of relativity, usually take the following approach: As a rule, before solving gravitational field equations, four restrictions (coordinate or gauge conditions) are imposed on the components of the Riemannian\footnote{The notational conventions employed here are those used by Landau and Lifshitz \cite{Landau-Lifshitz-1988}: Letters from the second half of the Latin alphabet, $m, n,...=0...3$ denote space-time indices. Greek letters $\alpha, \beta,...=1...3$ denote spatial indices. The metric $\gamma_{mn}$ is that of Minkowski space-time with $\gamma_{mn}={\rm diag}(+1,-1,-1,-1)$ in the Cartesian representation. The coordinates are formed such that $(ct,\vec{r})=(x^0,x^\alpha)$, where $c$ is the velocity of light. We employ the Einstein summation convention with indices being lowered or raised using $\gamma_{mn}$.} metric $g_{mn}$. These conditions extract a particular subset from an infinite set of space-time coordinates. Within this subset, the coordinates are linked by smooth differentiable transformations that do not change the coordinate conditions that were imposed. A set of differential coordinate conditions used in leading theories of relativistic reference systems, such as that recommended by the International Astronomical Union (see, for instance \cite{Soffel:2003cr}), are the harmonic gauge conditions.
The harmonic gauge has a very prominent role in gravitational physics, starting with the work of Fock \cite{Fock-book-1959}. In addition, a set of specific conditions designed to fix a particular reference frame is added to eliminate most of the remaining degrees of freedom, yielding an explicit form for the coordinate system associated with either inertial or accelerating frames.

In the present paper we introduce a new method for deriving the metric associated with the proper reference frame of an accelerated observer. To constrain the available degrees of freedom, we will also use the harmonic gauge. An argument in favor of choosing the harmonic gauge is that tremendous work in general relativity has been done with this gauge, which was found to be a simplifying and useful gauge for many kinds applications \cite{Soffel:2003cr}. Using the harmonic gauge, we develop the structure of both the direct and inverse coordinate transformations between inertial and accelerated reference frames. Such mutual transformations are not fully treated in the current literature, since usually either the direct \cite{DSX-I,Soffel:2003cr} or the inverse \cite{Brumberg-Kopejkin-1989-2} transformation is developed, but not both at the same time. The method presented in this paper could help to generalize contemporary theories of relativistic reference frames and to allow one to deal naturally with both transformations and relevant equations of motion in a unified formalism. Finally, the method presented here does not rely on a particular theory of gravitation; instead, it uses the covariant coordinate transformations to explore the dynamics in Minkowski space-time from a general perspective. Thus, any description from the standpoint of a metric theory of gravity must yield our results in the special relativistic limit.

We begin our discussion in Section \ref{sec:direct-tr} by generalizing the Lorentz transformations (\ref{eq:1.7a}) to the case of accelerated motion. This generalization can be done in the form of a slow motion approximation that admits expansion of all the quantities involved in the form of power series. This task is accomplished by introducing acceleration-dependent terms in the coordinate transformations. The resulting transformation is given in a general form that relies on a set of functions that are precisely determined in the subsequent sections.

The local coordinate system of an accelerated observer is not unique. We use the harmonic gauge to constrain the set of coordinates chosen in the accelerated reference frame. We carry out this task in Section \ref{sec:harm} and determine the metric tensor describing the accelerated reference frame and the structure of the coordinate transformations that satisfy the harmonic gauge conditions. We observe that the  metric in the accelerated frame has an elegant form that depends only on two harmonic potentials, which yield a powerful tool that allows reconstruction of the spatial part of the structure of the post-Galilean coordinate transformation functions between inertial and accelerating frames.

To fix the remaining degrees of freedom and to specify the proper reference frame of an accelerated observer, we introduce a set of dynamical conditions in Section \ref{sec:dyn}. Specifically, we require that the relativistic linear three-momentum of the accelerated observer in its proper reference frame to be conserved. This conservation leads to fixing the time-like coordinate in the accelerating frame, which allows to fix uniquely all the remaining terms.

A similar approach can also be carried out in reverse, establishing coordinate transformation rules from an accelerating to an inertial reference frame, which is done in Sec.~\ref{sec:inverse-tr}. Key to the approach presented in this section is the use of the contravariant metric for the accelerating frame, which leads in a straightforward manner to the inverse Jacobian matrix. This allows us to present the inverse transformations, in which the roles of the inertial and accelerating coordinates are reversed. The calculations are formally very similar to those presented in Secs.~\ref{sec:harm} and \ref{sec:dyn}, thus, to avoid repetitiveness, we show only the main results.

We conclude by discussing these results and presenting our recommendations for future research in Section~\ref{sec:end}. We also show the correspondence between our results and those obtained previously by other authors.

\section{Post-Galilean coordinate transformations and bookkeeping}
\label{sec:direct-tr}

To describe the dynamics of the $N$-body problem one usually introduces $N+1$ reference frames with their own coordinate charts. We need one {\it global} coordinate chart defined for the inertial reference frame that covers the entire system under consideration.  In the immediate vicinity of each of the $N$ bodies in the system we can also introduce a set of {\it local} coordinates defined in the frame associated with this body. In the remainder of this paper, we use $\{x^m\}$ to represent the coordinates of the global inertial frame and $\{y^m\}$ to be the local coordinates of the accelerated frame.

The most general post-Galilean form of a finite transformation between the coordinates of inertial $\{x^m\}$ and accelerating $\{y^m\}$ dynamically non-rotating reference frames, may be given in the following form \cite{Turyshev-96}:
{}
\begin{eqnarray}
x^0&=& y^0+c^{-2}{\cal K}(y^0, y^\epsilon)+
c^{-4}{\cal L}(y^0, y^\epsilon)+{\cal O}(c^{-6}),
\label{eq:trans-0}\\[3pt]
x^\mu&=& y^\mu+z^\mu_{0}(y^0)+
c^{-2}{\cal Q}^\mu(y^0, y^\epsilon)+{\cal O}(c^{-4}),
\label{eq:trans-a}
\end{eqnarray}
where $z^\mu_{0}$ is the Galilean vector connecting the spatial origins of two dynamically non-rotating frames, and we introduce the post-Galilean vector $x^\mu_{0}(y^0)$ connecting the spatial origins of the two frames
\begin{equation}
x^\mu_{0}(y^0)=z^\mu_{0}+c^{-2}{\cal Q}^\mu(y^0,0)+{\cal O}(c^{-4}).\label{eq:totvec}
\end{equation}
The functions ${\cal K}, {\cal L}$ and ${\cal Q}^\mu$ are yet unknown. It is anticipated that these functions depend only on the relative motion between the reference frames involved. Our objective is to construct an explicit form of the proper reference frame of the accelerating observer by determining the explicit functional form of the coordinate transformation functions ${\cal K}, {\cal L}$ and ${\cal Q}^\mu$. We shall refer to this approach as the ${\cal KLQ}$-formalism for relativistic reference frames.

In this formulation, we use the dimensioned parameter $c^{-1}$ as a bookkeeping device for order terms. For instance, when we write $c^{-2}{\cal K}$ in Eq.~(\ref{eq:trans-0}) above, this implies that the function ${\cal K}$ is of order $v^2y^0$ with $v\ll c$ being the characteristic velocity of an observer; hence, $c^{-2}{\cal K}$ is of order $(v/c)^2y^0$, which remains small relative to $y^0$. Similarly,  $c^{-4}{\cal L}\sim (v/c)^4y^0$ and $c^{-2}{\cal Q}^\mu \sim (v/c)^2 y^\mu$. Specifically, in the no-acceleration limit, Eqs.~(\ref{eq:trans-0})--(\ref{eq:trans-a}) should reduce to the Lorentz transformations given by Eqs.~(\ref{eq:1.7a}), which can be written in approximate form as:
{}
\begin{eqnarray}
x^0&=&\Big(1+{v^2_0\over2c^2} + {3v^4_0\over8c^4}\Big)y^0+
\Big(1 +{v^2_0\over2c^2}\Big){(\vec{v}_0\cdot\vec{y})\over c}
+{\cal O}\left(c^{-6}\right),
\label{eq:1.12a} \\
\vec{x}&=& \vec{y} +\Big(1 +{v^2_0\over2c^2}\Big)\vec{v}_0 c^{-1}y^0+{\vec{v}_0(\vec{v}_0\cdot\vec{y}) \over 2c^2}
+{\cal O}\left(c^{-4}\right),
\label{eq:1.12b}
\end{eqnarray}
corresponding to the following set of ${\cal KLQ}$ coordinate  transformation functions: ${\cal K}_0= \frac{1}{2}v^2_0y^0+c(\vec{v}_0\cdot\vec{y})+{\cal O}(c^{-4})$, ${\cal L}_0= \frac{3}{8}v^4_0y^0+\frac{1}{2}v^2_0c(\vec{v}_0\cdot\vec{y})+{\cal O}(c^{-2})$, ${\cal Q}^\alpha_0=\frac{1}{2}{\vec{v}_0(\vec{v}_0\cdot\vec{y})}+\frac{1}{2}{v^2_0}\vec{v}_0 c^{-1}y^0+{\cal O}(c^{-2})$, with $z_0^\mu={\vec v}_0c^{-1}y^0$. These approximations remain valid so long as $y^\mu/y^0\ll v_0/c$, a condition that is naturally satisfied along the world-line $y^\mu=0$.

The coordinate transformation rules for the general coordinate transformations above are easy to obtain and express in the form of the Jacobian matrix ${\partial x^m}/{\partial y^n}$. From Eqs.~(\ref{eq:trans-0})--(\ref{eq:trans-a}), we get:
{}
\begin{eqnarray}
{\partial x^0\over\partial y^0}
&=& 1 + c^{-2}{\partial {\cal K}\over\partial y^0}  +
c^{-4}{\partial {\cal L}\over\partial y^0}  + {\cal O}(c^{-6}),
 \label{eq:(C1a)}\qquad
{\partial x^0\over\partial y^\nu} =
c^{-2} {\partial {\cal K}\over\partial y^\nu}  +
c^{-4}{\partial {\cal L}\over\partial y^\nu}  +
{\cal O}(c^{-5}), \label{eq:(C1b)}\\
{}
{\partial x^\mu\over\partial y^0} &=& \frac{v^\mu_{0}}{c} +
c^{-2} {\partial {\cal Q}^\mu\over\partial y^0} + {\cal O}(c^{-5}), \label{eq:(C1c)}\qquad\qquad\quad\,
{\partial x^\mu\over \partial y^\nu}  =
\delta^\mu_\nu  +
 c^{-2}{\partial {\cal Q}^\mu\over\partial y^\nu}  + {\cal O}(c^{-4}), \label{eq:(C1d)}
\end{eqnarray}
where $v^\epsilon_0 =\dot z^\epsilon_0\equiv cd z^\epsilon_{0}/d y^0$ is the time-dependent relative velocity between the two frames.

The transformations given by Eqs. (\ref{eq:trans-0})--(\ref{eq:trans-a}) must be smooth and the corresponding Jacobian matrix ${\partial x^m}/{\partial y^n}$ must be non-singular, i.e., its determinant must be nonvanishing\footnote{We can find an upper limit for the distances at which the condition (\ref{(3.11b)}) remains valid. With the help of Eqs.~(\ref{eq:DeDoGa-L})--(\ref{eq:DeDoGa-Q}) and the solution for the function ${\cal K}$ given by (\ref{eq:fin-sol-K}) below, we can present Eq.~(\ref{(3.11b)}) as
$$
\hbox{det}
\Big(
{\partial  x^m\over \partial y^n}
\Big)
=1 - \frac{2}{c^2}\Big({\partial {\cal K}\over\partial y^0}  + {\textstyle\frac{1}{2}}
v^{}_{0\epsilon} v^\epsilon_0 \Big) +{\cal  O}(c^{-4})=
1+\frac{2}{c^2}\Big((a_\epsilon y^\epsilon)-\varphi_0\Big)+{\cal  O}(c^{-4}).
$$
Therefore, the transformations (\ref{eq:trans-0})--(\ref{eq:trans-a}) are non-singular up to the distances from the world-line that satisfy the following condition  $y\lesssim(c^2-2\varphi_0)/(2a_0)$. Note that, for any realistic scenario, this condition is not violated on solar system scales \cite{Kopejkin-1988}.}:
\begin{equation}
\hbox{det}
\Big(
{\partial  x^m\over \partial y^n}
\Big)
= 1 +  c^{-2}\Big({\partial {\cal K}\over\partial y^0}  +
v^{}_{0\epsilon} v^\epsilon_0 + {\partial {\cal Q}^\mu\over\partial y^\mu} \Big) +{\cal  O}(c^{-4})\not=0,
\label{(3.11b)}
\end{equation}
which guarantees the existence of inverse transformations, discussed in Sec.~\ref{sec:inverse-tr}.

To establish the form of the metric tensor in the moving frame, we apply the usual tensor transformation rule:
\begin{equation}
\eta_{mn}(y)=\frac{\partial x^k}{\partial y^m}\frac{\partial x^l}{\partial y^n}\gamma_{kl}(x(y)),
\label{eq:eta-loc_cov}
\end{equation}
which, together with Eqs.~(\ref{eq:(C1a)})--(\ref{eq:(C1d)}), allows us to compute the metric components in the arbitrarily moving frame:
{}
\begin{eqnarray}
\eta_{00}(y)&=&1+\frac{2}{c^2}\Big\{\frac{\partial {\cal K}}{\partial y^0}+{\textstyle\frac{1}{2}}v^{}_{0\epsilon} v_0^\epsilon\Big\}
+\frac{2}{c^4}\Big\{ \frac{\partial {\cal L}}{\partial y^0}+cv_{0\epsilon}\frac{\partial {\cal Q}^\epsilon_a}{\partial y^0}+
{\textstyle\frac{1}{2}}\Big(\frac{\partial {\cal K}}{\partial y^0}\Big)^2\Big\}+O(c^{-6}), \label{eq:eta00_cov}\\
{}
\eta_{0\alpha}(y)&=&\frac{1}{c}\Big\{
\frac{1}{c}{\partial {\cal K}\over\partial y^\alpha}+v_{0\alpha}\Big\} +
\frac{1}{c^3}\Big\{\frac{1}{c}\frac{\partial {\cal L}}{\partial y^\alpha}+c\gamma_{\alpha\epsilon}\frac{\partial {\cal Q}^\epsilon}{\partial y^0}+
v^{}_{0\lambda}\frac{\partial {\cal Q}^\lambda}{\partial y^\alpha}+\frac{1}{c}{\partial {\cal K}\over\partial y^\alpha}\frac{\partial {\cal K}}{\partial y^0}\Big\}+ {\cal O}(c^{-5}),
\label{eq:eta0a_cov}\\
{}
\eta_{\alpha\beta}(y)&=&
\gamma_{\alpha\beta}+\frac{1}{c^2}\Big\{
\frac{1}{c}{\partial {\cal K}\over\partial y^\alpha}
\frac{1}{c}{\partial {\cal K}\over\partial y^\beta}
+\gamma_{\alpha\lambda}\frac{\partial {\cal Q}^\lambda}{\partial y^\beta}+\gamma_{\beta\lambda}\frac{\partial {\cal Q}^\lambda}{\partial y^\alpha}\Big\}+O(c^{-4}). \label{eq:etaab_cov}
\end{eqnarray}

Eqs.~(\ref{eq:eta00_cov})--(\ref{eq:etaab_cov}) show the structure of the metric tensor of the Minkowski space-time in an arbitrarily moving reference frame.  The actual dependence of $\eta_{mn}$ on the transformation functions  (${\cal K},{\cal L},{\cal Q}^\alpha$) will be important to define the proper relativistic reference frame of an accelerated observer under the harmonic gauge. This will be investigated next.

\section{Imposing the harmonic gauge conditions}
\label{sec:harm}

The dynamical condition, i.e., the requirement that the spatial origin of the transformed system of coordinates $x^m=f(y^m)$ is to move along a specific world-line, does not uniquely define $y^m$. The existence of this coordinate freedom allows us to impose the harmonic (de Donder) gauge condition in the local frame\footnote{In Secs.~\ref{sec:harm} and \ref{sec:dyn}, when convenient and unambiguous, we shall use the abbreviated form of the partial derivative operator to represent partial derivatives with respect to $\{y^m\}$: $\partial_m=\partial/\partial y^m$.}:
\begin{equation}
\partial_m\big({\sqrt{-\eta}}\eta^{mn}\big) = 0.
\label{eq:(DeDg)OK}
\end{equation}
The vanishing of the covariant derivative of the metric tensor
allows us to present Eq.~(\ref{eq:(DeDg)OK}) in the following equivalent form:
$\eta^{kl}\Gamma^n_{kl}(\eta)=0.$
Remembering the transformation rules of the Christoffel symbols under general coordinate transformations \cite{Landau-Lifshitz-1988}, one can verify that the latter equation is equivalent to imposing the harmonic conditions on the transformation functions in Eqs.~(\ref{eq:trans-0})--(\ref{eq:trans-a}):
{}
\begin{equation}
\Box_\eta x^m= 0,
\label{eq:(DeDg)x0-box}
\end{equation}
where $\Box_\eta=(\sqrt{-\eta})^{-1}\partial_m(\sqrt{-\eta}\eta^{mn}\partial _n)$ denotes the covariant d'Alembertian, in this case acting on $x^m$, which are treated as individual scalar functions.

Therefore, on the one hand the harmonic gauge imposes restrictions on the partial derivatives of the metric tensor, as seen in Eq.~(\ref{eq:(DeDg)OK}). On the other hand, it restricts the choice of admissible coordinate transformations only to those that satisfy the harmonic equation (\ref{eq:(DeDg)x0-box}). These two consequences of imposing the harmonic gauge will be used to establish the structure that the metric tensor (\ref{eq:eta00_cov})--(\ref{eq:etaab_cov}) must satisfy under the harmonic coordinate transformations (\ref{eq:trans-0})--(\ref{eq:trans-a}) and also to constrain the form of the transformation functions (${\cal K},{\cal L},{\cal Q}^\alpha$).

\subsection{The form of the metric tensor in a moving frame}
\label{sec:harm-metric}

We require the proper reference frame of a moving observer to exhibit no rotation or shear of its coordinate axes. The components of the metric tensor $\eta_{mn}$ that satisfy these requirements under the harmonic gauge conditions (\ref{eq:(DeDg)OK}) must, therefore, satisfy the following set of partial differential equations
{}
\begin{eqnarray}
\eta^{[1]}_{0\alpha}&=&{\cal O}(c^{-4}),
 \label{eq:(DeDgGa)OKs01+}\\
\frac{1}{2}c\partial_0 \Big\{\eta^{[2]}_{00}
-\gamma^{\epsilon\lambda}\eta^{[2]}_{\epsilon\lambda}\Big\} +\partial^\nu \eta^{[3]}_{0\nu} &=& {\cal O}(c^{-2}), \hskip 20pt
 \label{eq:(DeDgGa)OKs0+}\\
\partial^\beta \Big\{\eta^{[2]}_{\alpha\beta}
-{1\over2} \gamma_{\alpha\beta}\Big(\eta^{[2]}_{00} +\gamma^{\epsilon\lambda}\eta^{[2]}_{\epsilon\lambda}\Big)\Big\} &=& {\cal O}(c^{-2}).
 \label{eq:(DeDgGa)OKsa+}
\end{eqnarray}

By formally integrating Eq.~(\ref{eq:(DeDgGa)OKsa+}) and choosing the solution corresponding to a subset of the harmonic gauge with spatially isotropic coordinates, we are led to the following form of the gauge conditions (\ref{eq:(DeDgGa)OKs01+})--(\ref{eq:(DeDgGa)OKsa+}):
{}
\begin{eqnarray}
\eta^{[1]}_{0\alpha}&=&{\cal O}(c^{-4}),
 \label{eq:(DeDgGa)OKs01}\\
2c\partial_0 \eta^{[2]}_{00} +\partial^\nu \eta^{[3]}_{0\nu} &=&
{\cal O}(c^{-2}), \hskip 20pt
 \label{eq:(DeDgGa)OKs0}\\
\eta^{[2]}_{\alpha\beta}+\gamma_{\alpha\beta}\eta^{[2]}_{00}&=& {\cal O}(c^{-2}).
 \label{eq:(DeDgGa)OKsa}
\end{eqnarray}

Eqs.~(\ref{eq:(DeDgGa)OKs01})--(\ref{eq:(DeDgGa)OKsa}) represent a set of harmonic conditions on the metric tensor $\eta_{mn}$ given by Eqs.~(\ref{eq:eta00_cov})--(\ref{eq:etaab_cov}) in the coordinates of a moving reference frame. This set of gauge conditions forms the foundation of our method of constructing a proper reference frame of an arbitrarily moving observer. As a result, the metric representing space-time in the moving frame may be  presented in the following elegant isotropic form that depends only on two harmonic potentials:
\begin{eqnarray}
\eta_{00}(y)&=& 1-\frac{2}{c^2}u(y)+\frac{2}{c^4}u^2(y)+O(c^{-6}),
\label{eq:eta_00-cov}\\
\eta_{0\alpha}(y)&=& -\gamma_{\alpha\lambda}\frac{4}{c^3}u^\lambda(y)+O(c^{-5}),
\label{eq:eta_0a-cov}\\
\eta_{\alpha\beta}(y)&=& \gamma_{\alpha\beta}+\gamma_{\alpha\beta}\frac{2}{c^2} u(y)+O(c^{-4}),
\label{eq:eta_ab-cov}
\end{eqnarray}
where we introduced the scalar $u(y)$ and the vector $u^\alpha(y)$ potentials, defined as:
{}
\begin{eqnarray}
u(y)&=&-\frac{\partial {\cal K}}{\partial y^0}-
{\textstyle\frac{1}{2}}v_{0}{}_\epsilon v_{0}^\epsilon -
\frac{1}{c^2}\Big\{ \frac{\partial {\cal L}}{\partial y^0}+cv_{0\epsilon}\frac{\partial {\cal Q}^\epsilon}{\partial y^0}+
{\textstyle\frac{1}{2}}\Big(\frac{\partial {\cal K}}{\partial y^0}\Big)^2-\Big(\frac{\partial {\cal K}}{\partial y^0}+
{\textstyle\frac{1}{2}}v_{0}{}_\epsilon v_{0}^\epsilon\Big)^2\Big\}+O(c^{-4}),
\label{eq:pot_loc-w_0-cov}\\
u^\alpha(y)&=& -
{\textstyle\frac{1}{4}}\Big\{ \gamma^{\alpha\epsilon}\frac{1}{c}\frac{\partial {\cal L}}{\partial y^\epsilon}+\gamma^{\alpha\epsilon}v_{0\lambda}\frac{\partial {\cal Q}^\lambda}{\partial y^\epsilon}+c\frac{\partial {\cal Q}^\alpha}{\partial y^0}-v_{0}^\alpha\frac{\partial {\cal K}}{\partial y^0}\Big\}+O(c^{-2}).
\label{eq:pot_loc-w_a-cov}
\end{eqnarray}
It follows from Eq.~(\ref{eq:(DeDgGa)OKs0}) that these potentials satisfy the continuity equation:
{}
\begin{equation}
c\partial_0 u+ \partial_\epsilon u^\epsilon=
{\cal O}(c^{-2}).  \label{eq:(DeDgGa)cont-eq}
\end{equation}

With the help of Eqs.~(\ref{eq:form-inv_0a})--(\ref{eq:DeDoGa-Q}), derived in Sec.~\ref{seq:harmform} below, we can verify that the potentials $u$ and $u^\alpha$ satisfy the following harmonic conditions:
{}
\begin{equation}
\Box_y u={\cal O}(c^{-4}), \qquad
\Delta_y  u^\alpha = {\cal O}(c^{-2}),  \label{eq:w-Dalemb}
\end{equation}
where $\Box_y =\gamma^{mn}\partial_m\partial_n=\partial_0^2+\Delta_y$ and $\Delta_y =\gamma^{\epsilon\lambda}\partial_\epsilon\partial_\lambda$ are the flat-space d'Alembert and Laplace operators in local $\{y^m\}$ coordinates.\footnote{An interesting consequence of the solution for the potentials $u$ and $u^\alpha$ is the fact that this choice ensures that the induced inertial space-time is flat, and thus, the Ricci tensor of this space-time vanishes, $R_{mn}(\eta)=0$. This can be verified by a direct calculation of the Ricci tensor $R_{mn}$ with the metric $\eta_{mn}$ given by Eqs.~(\ref{eq:eta_00-cov})--(\ref{eq:eta_ab-cov}) and also by applying Eqs.~(\ref{eq:w-Dalemb}):
\begin{eqnarray*}
R_{00}&=&\Box_y \Big\{c^{-2}\eta_{00}^{[2]}+c^{-4}\Big(\eta_{00}^{[4]}-{\textstyle\frac{1}{2}}(\eta_{00}^{[2]})^2\Big)+{\cal O}(c^{-6})\Big\}=-\frac{2}{c^2}\Box_y \Big\{u+{\cal O}(c^{-4})\Big\}={\cal O}(c^{-6}),
 \label{eq:(DeDgGa)in-01}\\
R_{0\alpha}&=&\Delta_y \Big\{c^{-3}\eta_{0\alpha}^{[3]}+{\cal O}(c^{-5})\Big\}=-\gamma_{\alpha\lambda}\frac{4}{c^3}\Delta_y \Big\{u^\lambda+{\cal O}(c^{-2})\Big\}={\cal O}(c^{-5}),
 \label{eq:(DeDgGa)in-02}\\
R_{\alpha\beta}&=&\Delta_y \Big\{c^{-2}\eta_{\alpha\beta}^{[2]}+{\cal O}(c^{-4})\Big\}=\gamma_{\alpha\beta}\frac{2}{c^2} \Delta_y \Big\{u+{\cal O}(c^{-2})\Big\}={\cal O}(c^{-4}).
 \label{eq:(DeDgGa)in-03}
\end{eqnarray*}
This observation regarding the potentials $u$ and $u^\alpha$ will be important in the case of gravitational dynamics of $N$-body systems, which will be studied in a subsequent paper.
}
We shall use the term ``harmonic metric tensor'' to describe the metric tensor (\ref{eq:eta_00-cov})--(\ref{eq:eta_ab-cov}), expressed in terms of the harmonic potentials $u$ and $u^\alpha$, given by Eqs.~(\ref{eq:pot_loc-w_0-cov}) and (\ref{eq:pot_loc-w_a-cov}), which satisfy Eqs.~(\ref{eq:(DeDgGa)cont-eq}) and (\ref{eq:w-Dalemb}).

\subsection{The form of the functions of the harmonic coordinate transformations}
\label{seq:harmform}

As discussed in Sec.~\ref{sec:harm-metric}, Eqs.~(\ref{eq:(DeDgGa)OKs01}) and (\ref{eq:(DeDgGa)OKsa}) provide valuable constraints on the form of the metric tensor in a moving frame. As a matter of fact, these equations provide two additional restrictions on ${\cal K}$ and ${\cal Q}^\alpha$. It follows from Eqs.~(\ref{eq:(DeDgGa)OKs01}), (\ref{eq:(DeDgGa)OKsa}) and the form of the metric tensor $\eta_{mn}$ given by Eqs.~(\ref{eq:eta00_cov})--(\ref{eq:etaab_cov}) that these two functions must also satisfy two first order partial differential equations:
{}
\begin{eqnarray}
\frac{1}{c}{\partial {\cal K}\over\partial y^\alpha} +v_{0\alpha}
&=&{\cal O}(c^{-4}),
\label{eq:form-inv_0a} \\
\frac{1}{c}{\partial {\cal K}\over\partial y^\alpha}
\frac{1}{c}{\partial {\cal K}\over\partial y^\beta}+\gamma_{\alpha\lambda}\frac{\partial {\cal Q}^\lambda}{\partial y^\beta}+\gamma_{\beta\lambda}\frac{\partial {\cal Q}^\lambda}{\partial y^\alpha}+2\gamma_{\alpha\beta}\Big(\frac{\partial {\cal K}}{\partial y^0}+
{\textstyle\frac{1}{2}}v^{}_{0\epsilon} v_{0}^\epsilon\Big) &=& {\cal O}(c^{-2}).
\label{eq:form-inv_ab}
\end{eqnarray}

We now explore the alternative form of the harmonic gauge given by Eq.~(\ref{eq:(DeDg)x0-box}). Substituting the coordinate transformations (\ref{eq:trans-0})--(\ref{eq:trans-a}) into Eq.~(\ref{eq:(DeDg)x0-box}), we can see that the harmonic gauge conditions restrict the coordinate transformation functions ${\cal K}, {\cal L}$ and ${\cal Q}^\alpha$ such that they must satisfy the following set of second order partial differential equations:
{}
\begin{eqnarray}
\gamma^{\epsilon\lambda}\frac{\partial^2 {\cal K}}{\partial y^\epsilon \partial y^\lambda}
&=& {\cal O}(c^{-4}), \label{eq:DeDoGa-K}\\
c^2\frac{\partial^2 {\cal K}}{\partial {y^0}^2}+
\gamma^{\epsilon\lambda}\frac{\partial^2 {\cal L}}{\partial y^\epsilon \partial y^\lambda}
&=& {\cal O}(c^{-2}), \label{eq:DeDoGa-L}\\
a_{0}^\alpha+
\gamma^{\epsilon\lambda}\frac{\partial^2 {\cal Q}^\alpha}{\partial y^\epsilon \partial y^\lambda}
&=& {\cal O}(c^{-2})
\label{eq:DeDoGa-Q}.
\end{eqnarray}

The general solution to these elliptic-type equations for the functions ${\cal K}, {\cal L}$ and ${\cal Q}^\alpha$ in Eqs.~(\ref{eq:DeDoGa-K})--(\ref{eq:DeDoGa-Q}) consist of two parts: a fundamental solution of the homogeneous Laplace equation and a particular solution of the inhomogeneous Poisson equation (except for Eq.~(\ref{eq:DeDoGa-K}), which is homogeneous). These solutions can be written in the form of a Taylor series expansion in terms of irreducible Cartesian tensors, which are
symmetric and trace-free (STF) \cite{Thorne-1980}.

The two sets of partial differential equations for ${\cal K}, {\cal L}$ and ${\cal Q}^\alpha$ given by Eqs.~(\ref{eq:form-inv_0a})--(\ref{eq:form-inv_ab}) and (\ref{eq:DeDoGa-K})--(\ref{eq:DeDoGa-Q}) can be used to determine the general structure of these functions.

\subsubsection{Determining the structure of ${\cal K}$}

The general solution to Eq.~(\ref{eq:DeDoGa-K}) with regular behavior on the world-line (i.e., omitting terms divergent when $|\vec y|\rightarrow 0$ or solutions not differentiable at $|\vec{y}|=0$) can be given in the following form:
\begin{equation}
{\cal K} (y) = \kappa_{0}+ \kappa_{0\mu} y^\mu + \delta \kappa+ {\cal O}(c^{-4}), \qquad {\rm where} \qquad
\delta \kappa=\sum_{k\ge 2}\frac{1}{k!}\kappa_{\mu_1...\mu_k} (y^0)y^{\mu_1...}y^{\mu_k} + {\cal O}(c^{-4}),
\label{eq:K-gen}
\end{equation}
with $\kappa_{\mu_1...\mu_k} (y^0)$ being STF tensors \cite{Thorne-1980}, which depend only on the timelike coordinate $y^0$. Substituting this form of the function ${\cal K}$ into equation (\ref{eq:form-inv_0a}), we the find solutions for $\kappa_{0\mu}$ and $\kappa_{\mu_1...\mu_k}$:
{}
\begin{equation}
\kappa_{0\mu} =- c v_{0\mu}  + {\cal O}(c^{-4}),\quad
\kappa_{\mu_1...\mu_k}={\cal O}(c^{-4}), \quad k\ge 2.
\label{eq:Ka-cov0}
\end{equation}

As a result, the function ${\cal K}$ that satisfies the harmonic gauge conditions is determined to be
{}
\begin{equation}
{\cal K} (y) = \kappa_{0} - c (v_{0\mu}y^\mu)  + {\cal O}(c^{-4}).
\label{eq:Ka-cov}
\end{equation}
This expression completely fixes the spatial dependence of the function ${\cal K}$, but still has an unknown dependence on the timelike coordinate via the function $\kappa_{0} (y^0)$.

\subsubsection{Determining the structure of ${\cal Q}^\alpha$}

The general solution for the function ${\cal Q}^\alpha$ that satisfies Eq.~(\ref{eq:DeDoGa-Q}) may be presented as a sum of a solution of the inhomogeneous Poisson equation and a solution of the homogeneous Laplace equation. Furthermore, the part of that solution with regular behavior in the vicinity of the world-line may be given in the following form:
\begin{equation}
{\cal Q}^\alpha (y) = q^\alpha_{0}+ q^\alpha_{0\mu} y^\mu +
{\textstyle\frac{1}{2}}q^\alpha_{0\mu\nu} y^\mu y^\nu + \delta\xi^\alpha+ {\cal O}(c^{-2}),
\label{eq:K-gen+}
\end{equation}
where $q^\alpha_{0\mu\nu}$ can be determined directly from Eq.~(\ref{eq:DeDoGa-Q}) and the function $\delta\xi^\alpha$ satisfies the Laplace equation
\begin{eqnarray}
\Delta_y \delta\xi^\alpha &=& {\cal O}(c^{-2}).
\label{eq:DeDoGa-a_xi+}
\end{eqnarray}

We can see that Eq.~(\ref{eq:DeDoGa-Q}) can be used to determine $q^\alpha_{0\mu\nu}$, but would leave the other terms in the equation unspecified. To determine these terms, we use Eq.~(\ref{eq:form-inv_ab}) together with Eq.~(\ref{eq:form-inv_0a}), and get:
{}
\begin{eqnarray}
v^{}_{0\alpha} v^{}_{0\beta}+\gamma_{\alpha\lambda}\frac{\partial {\cal Q}^\lambda}{\partial y^\beta}+\gamma_{\beta\lambda}\frac{\partial {\cal Q}^\lambda}{\partial y^\alpha}+2\gamma_{\alpha\beta}\Big(\frac{\partial {\cal K}}{\partial y^0}+
{\textstyle\frac{1}{2}}v_{0}{}_\epsilon v_{0}^\epsilon\Big)&=&{\cal O}(c^{-2}).
\label{eq:Q-form-inv-cov}
\end{eqnarray}

Using the intermediate solution (\ref{eq:Ka-cov}) for the function ${\cal K}$ in Eq.~(\ref{eq:Q-form-inv-cov}), we obtain the following equation for ${\cal Q}^\alpha$:
{}
\begin{eqnarray}
v^{}_{0\alpha} v^{}_{0\beta}+\gamma_{\alpha\lambda}\frac{\partial {\cal Q}^\lambda}{\partial y^\beta}+\gamma_{\beta\lambda}\frac{\partial {\cal Q}^\lambda}{\partial y^\alpha}+2\gamma_{\alpha\beta}\Big(\frac{\partial \kappa_0}{\partial y^0}+
{\textstyle\frac{1}{2}}v_{0}{}_\epsilon v_{0}^\epsilon- a^{}_{0\epsilon} y^\epsilon\Big)&=&{\cal O}(c^{-2}).
\label{eq:form-inv_exp-cov}
\end{eqnarray}

A trial solution to Eq.~(\ref{eq:form-inv_exp-cov}) may be given in the following general from:
{}
\begin{equation}
{\cal Q}^\alpha = q^{\alpha}_{0}+
c_1 v_{0}^\alpha v^{}_{0\epsilon} y^{\epsilon} +
c_2 v^{}_{0\epsilon} v_{0}^\epsilon y^{\alpha} +
c_3 a^{\alpha}_{0} y^{}_{\epsilon} y^\epsilon+
c_4 a^{}_{0\epsilon} y^\epsilon y^{\alpha} +
c_5\big(\frac{\partial \kappa_0}{\partial y^0}+
{\textstyle\frac{1}{2}}v_{0}{}_\epsilon v_{0}^\epsilon\big)y^\alpha
+y^{}_\epsilon\omega_{0}^{\epsilon\alpha}+\delta\xi^\mu(y),
\label{eq:delta-Q-c}
\end{equation}
where $q^{\alpha}_{0}$ and the antisymmetric matrix $\omega_{0}^{\alpha\epsilon}=-\omega_{0}^{\epsilon\alpha}$ are functions of the timelike coordinate $y^0$; $c_1,..., c_5$ are constants; and $\delta\xi^\mu(y)$, given by Eq.~(\ref{eq:DeDoGa-a_xi+}), is at least of third order in the spatial coordinates $y^\mu$, namely $\delta\xi^\mu(y)\propto{\cal O}(|y^\mu|^3)$. Direct substitution of Eq.~(\ref{eq:delta-Q-c}) into Eq.~(\ref{eq:form-inv_exp-cov}) results in the following unique solution for these coefficients:
{}
\begin{equation}
c_1 = -\frac{1}{2},\quad
c_2 = 0,\quad
c_3 = -\frac{1}{2}, \quad
c_4 = 1,\quad
c_5 = -1.
\label{eq:delta-Q-c-sol}
\end{equation}
As a result, the function ${\cal Q}^\alpha$ has the following structure
\begin{eqnarray}
{\cal Q}^\alpha(y)&=& q^{\alpha}_{0} -
\Big({\textstyle\frac{1}{2}}v^\alpha_{0} v^\epsilon_{0}+
\omega_{0}^{\alpha\epsilon}+\gamma^{\alpha\epsilon}\big(\frac{\partial \kappa_0}{\partial y^0}+
{\textstyle\frac{1}{2}}v_{0}{}_\lambda v_{0}^\lambda\big)\Big){y}_\epsilon +
a^{}_{0\epsilon}\Big(y^\alpha y^\epsilon-
{\textstyle\frac{1}{2}}\gamma^{\alpha\epsilon}{y}_\lambda y^\lambda\Big)
+\delta\xi^\alpha(y),
\label{eq:d-Q+}
\end{eqnarray}
where $q^{\alpha}_{0}$ and $\omega_{0}^{\alpha\epsilon}$ are yet to be determined.

By substituting (\ref{eq:d-Q+}) into (\ref{eq:form-inv_exp-cov}), we see that the function $\delta\xi^\alpha(y)$ in Eq.~(\ref{eq:d-Q+}) must satisfy the equation:
{}
\begin{equation}
\partial^\alpha \delta\xi^\beta + \partial^\beta \delta\xi^\alpha ={\cal O}(c^{-2}).
\label{eq:form-inv_xi-cov}
\end{equation}
We keep in mind that the function $\delta\xi^\alpha(y)$ must also satisfy Eq.~(\ref{eq:DeDoGa-a_xi+}). The solution to the partial differential equation (\ref{eq:DeDoGa-a_xi+}) with regular behavior on the world-line (i.e., when $|\vec y|\rightarrow 0$) can be given in powers of $y^\mu$ as
\begin{equation}
\delta\xi^\alpha(y)=\sum_{k\ge 3}\frac{1}{k!}\delta\xi^\alpha_{0\,\mu_1...\mu_k}(y^0)y^{\mu_1}{}^{...}y^{\mu_k}+{\cal O}(|y^{\mu}|^K) + {\cal O}(c^{-2}),
\label{eq:delxi}
\end{equation}
where $\delta\xi^\alpha_{0\,\mu_1...\mu_k} (y^0)$ being STF tensors that depend only on timelike coordinate. Using the solution (\ref{eq:delxi}) in Eq.~(\ref{eq:form-inv_xi-cov}), we can see that $\delta\xi^\alpha_{0\,\mu_1...\mu_k}$ is also antisymmetric with respect to the index $\alpha$ and any of the spatial indices $\mu_1...\mu_k$. Combination of these two conditions suggests that $\delta\xi^\alpha_{0\,\mu_1...\mu_k}=0$ for all $k\ge 3$, thus
{}
\begin{equation}
\delta\xi^\alpha(y)=0.
\end{equation}

Therefore, application of the harmonic gauge conditions leads to the following form of the function ${\cal Q}^\alpha$:
\begin{eqnarray}
{\cal Q}^\alpha(y)&=& q^{\alpha}_{0} -
\Big({\textstyle\frac{1}{2}}v^\alpha_{0} v^\epsilon_{0}+
\omega_{0}^{\alpha\epsilon}+\gamma^{\alpha\epsilon}\big(\frac{\partial \kappa_0}{\partial y^0}+
{\textstyle\frac{1}{2}}v_{0}{}_\lambda v_{0}^\lambda\big)\Big){y}_\epsilon +
a^{}_{0\epsilon}\Big(y^\alpha y^\epsilon-
{\textstyle\frac{1}{2}}\gamma^{\alpha\epsilon}{y}_\lambda y^\lambda\Big)+ {\cal O}(c^{-2}),
\label{eq:d-Q}
\end{eqnarray}
where $q^{\alpha}_{0},\omega_{0}^{\alpha\epsilon}$ and $\kappa_0$  are yet to be determined.

\subsubsection{Determining the structure of ${\cal L}$}

We now turn our attention to the second gauge condition on the temporal coordinate transformation, Eq.~(\ref{eq:DeDoGa-L}). Using the intermediate solution (\ref{eq:Ka-cov}) for the function ${\cal K}$, we obtain the following equation for ${\cal L}$:
{}
\begin{eqnarray}
\gamma^{\epsilon\lambda}\frac{\partial^2 {\cal L}}{\partial y^\epsilon \partial y^\lambda}
&=& -c^2\frac{\partial^2 {\cal K}}{\partial {y^0}^2}+
{\cal O}(c^{-2})=c\big(v^{}_{0\epsilon} a_{0}^\epsilon+
\dot a^{}_{0\epsilon} y^\epsilon\big)-c^2\frac{\partial}{\partial y^0}\Big(\frac{\partial \kappa_0}{\partial y^0}+
{\textstyle\frac{1}{2}}v_{0}{}_\epsilon v_{0}^\epsilon\Big)+{\cal O}(c^{-2}).
\label{eq:DeDoGa-0_LK}
\end{eqnarray}

The general solution of Eq.~(\ref{eq:DeDoGa-0_LK}) for ${\cal L}$  may be presented as a sum of a solution $\delta {\cal L}$ for the inhomogeneous Poisson equation and a solution $\delta{\cal L}_0$ of the homogeneous Laplace equation.
A trial solution of the  inhomogeneous equation to this equation, $\delta{\cal L}$, is sought in the following form:
{}
\begin{equation}
\delta{\cal L} =
ck_1 (v^{}_{0\epsilon} a_{0}^\epsilon) (y^{}_{\mu} y^\mu)+
ck_2 (\dot{a}_{0\epsilon}y^\epsilon)(y_\nu y^\nu)-
k_3c^2\frac{\partial}{\partial y^0}\Big(\frac{\partial \kappa_0}{\partial y^0}+
{\textstyle\frac{1}{2}}v_{0}{}_\epsilon v_{0}^\epsilon\Big)(y_\nu y^\nu) +{\cal O}(c^{-2}),
\label{eq:delta-L}
\end{equation}
where $k_1, k_2, k_3$ are some constants. Direct substitution of (\ref{eq:delta-L}) into (\ref{eq:DeDoGa-0_LK}) yields the following values for the coefficients:
{}
\begin{equation}
k_1 = \frac{1}{6},\quad
k_2 = \frac{1}{10}, \quad
k_3 = \frac{1}{6}.
\label{eq:delta-Q-k-sol}
\end{equation}
As a result, the solution for $\delta{\cal L}$ that satisfies the harmonic gauge conditions has the following form:
{}
\begin{eqnarray}
\delta{\cal L}(y)&=&
{\textstyle\frac{1}{6}}c
\big(v^{}_{0\epsilon} a_{0}^\epsilon\big)(y^{}_\nu y^\nu)+
{\textstyle\frac{1}{10}}c(\dot{a}_{0\epsilon}y^\epsilon)
(y^{}_\nu y^\nu) -
{\textstyle\frac{1}{6}}c^2\frac{\partial}{\partial y^0}\Big(\frac{\partial \kappa_0}{\partial y^0}+
{\textstyle\frac{1}{2}}v_{0}{}_\epsilon v_{0}^\epsilon\Big)(y_\nu y^\nu) +{\cal O}(c^{-2}).
\end{eqnarray}

The solution for the homogeneous equation (\ref{eq:DeDoGa-0_LK}) with regular behavior on the world-line (i.e., when $|\vec{y}|\rightarrow 0$) may be presented as follows:
{}
\begin{eqnarray}
{\cal L}_0(y)&=&\ell_{0}(y^0)+\ell_{0\lambda}(y^0)\,y^\lambda+
{\textstyle\frac{1}{2}}\ell_{0\lambda\mu}(y^0)\,y^\lambda y^\mu
+\delta\ell(y),
\end{eqnarray}
where $\ell_{0\lambda\mu}$ is an STF tensor of second rank and $\delta\ell$ is a function formed from similar STF tensors of higher order:
\begin{equation}
\delta\ell(y)=\sum_{k\geq3}\frac{1}{k!}\delta\ell_{0\,\mu_1...\mu_k}(y^0)y^{\mu_1}{}^{...}y^{\mu_k}+{\cal O}(|y^{\mu}|^K).
\end{equation}

Finally, the general solution of Eq.~(\ref{eq:DeDoGa-0_LK}) may be
presented as a sum of the special solution $\delta{\cal L}$ of the inhomogeneous equation and the solution ${\cal L}_0$ of the homogeneous equation $\Delta_y {\cal L}=0$.
Therefore, the general solution for the gauge equations for the function ${\cal L}(y)= {\cal L}_0+\delta{\cal L}$ has the following form:
{}
\begin{eqnarray}
{\cal L}(y)&=&\ell_{0}+\ell_{0\lambda}\,y^\lambda+
{\textstyle\frac{1}{2}}\ell_{0\lambda\mu}\,y^\lambda y^\mu
+\delta\ell(y)+\nonumber\\[3pt]
&+&{\textstyle\frac{1}{6}}c
\big(v^{}_{0\epsilon} a_{0}^\epsilon\big)(y^{}_\nu y^\nu)+
{\textstyle\frac{1}{10}}c(\dot{a}_{0\epsilon}y^\epsilon)
(y^{}_\nu y^\nu) -
{\textstyle\frac{1}{6}}c^2\frac{\partial}{\partial y^0}\Big(\frac{\partial \kappa_0}{\partial y^0}+
{\textstyle\frac{1}{2}}v_{0}{}_\epsilon v_{0}^\epsilon\Big)(y_\nu y^\nu) +{\cal O}(c^{-2}).
\label{eq:L-gen-cov}
\end{eqnarray}

We have determined the structure of the transformation functions ${\cal K}, {\cal L}$, and ${\cal Q}^\alpha$, which is imposed by the harmonic gauge. Specifically, the harmonic structure for ${\cal K}$ is given by Eq.~(\ref{eq:Ka-cov}), the function  ${\cal Q}^\alpha$ was determined to have the structure given by Eq.~(\ref{eq:DeDoGa-0_LK}), and the structure for ${\cal L}$ is given by  Eq.~(\ref{eq:L-gen-cov}). Note that the harmonic gauge conditions allow us to reconstruct the structure of the functions only with respect to spatial coordinates $y^\mu$. The time-dependent functions $\kappa_0, q^{\alpha}_{0},\omega_{0}^{\alpha\epsilon}$, $\ell_{0}, \ell_{0\lambda}, \ell_{0\lambda\mu}$, and $\delta\ell_{0\,\mu_1...\mu_k}$ cannot be determined from the gauge conditions alone. We need to apply another set of conditions that would dynamically define the proper reference frame of a arbitrarily moving observer, thereby  fixing these time-dependent functions. This procedure will be discussed in the following section.

\section{Dynamical conditions for a proper reference frame}
\label{sec:dyn}

An accelerating observer that remains at rest with respect to an accelerating frame does so because of the balance between an external (physical) force that causes the observer to accelerate and the fictitious frame-reaction force that exists due to the choice of accelerating coordinates. The effects of both forces can be modeled in the form of an appropriately chosen effective metric $\eta^{\rm eff}_{mn}$ in the accelerating frame. In the case of complete balance, there will be no net force acting on the observer with respect to this effective metric, allowing it to be at rest in what we shall call its proper reference frame. The motion of the observer in this frame, then, will resemble a free fall that follows a geodesic with respect to the metric $\eta^{\rm eff}_{mn}$. Thus, the observer's ordinary relativistic linear three-momentum, calculated in the accelerating frame and with respect to $\eta^{\rm eff}_{mn}$, should be conserved. We can explore these conditions by writing down the Lagrangian of a test particle that represents the observer. By imposing further gauge conditions on the metric $\eta^{\rm eff}_{mn}$, we find that it is possible to eliminate all the remaining unknown components of ${\cal K}$, ${\cal L}$ and ${\cal Q}^\alpha$.

\subsection{Representing an external force using a fictitious harmonic metric}
\label{sec:fict-metr}

We imagine an observer that remains at rest in the accelerating frame $\{y^m\}$. We assume that this observer's acceleration $b^\alpha$ is due to an external universal scalar potential that, in the vicinity of the observer's world-line, admits the following representation:
{}
\begin{equation}
\varphi(y)=\varphi_0-(b_\epsilon y^\epsilon),
\label{eq:scal-pot}
\end{equation}
where $\varphi_0=\varphi_0(y^0)$ is the external background potential on the observer's world-line and $b^\alpha(y^0)=-\partial^\alpha\varphi$ is the external acceleration acting on the observer.

We incorporate the external potential into an effective metric, expressed in terms of ${y^m}$. This allows us to use a single metric to capture all physical processes occurring near the world-line: the external potential, $\varphi(y)$, that exerts the force accelerating the observer, and the frame potentials $u$ and $u^\alpha$ that generate a fictitious frame-reaction force that balances the force due to the external potential, keeping the observer at rest with respect to the accelerating frame. We shall denote this effective metric $\eta^{\rm eff}_{mn}$. At the Newtonian order (or up to ${\cal O}(c^{-4})$ in the $\eta^{\rm eff}_{00}$ component of the metric), we can write $\eta^{\rm eff}_{mn}$ by modifying the metric (\ref{eq:eta_00-cov})--(\ref{eq:eta_ab-cov}) as follows:
\begin{equation}
ds^2=\Big(1-\frac{2}{c^2}\big(u(y)+\varphi(y)\big)+
{\cal O}(c^{-4})\Big)(dy^0)^2+{\cal O}(c^{-3})dy^0dy^\lambda+
\big(\gamma_{\epsilon\lambda}+{\cal O}(c^{-2})\big)dy^\epsilon dy^\lambda.
\label{eq:inert-metr-01}
\end{equation}
In this metric, the potential $\varphi$ is responsible for the external force and $u$ is the inertial frame-reaction potential characterizing the accelerating reference frame, as introduced in the previous section and yet unknown. If the forces produced by the two potentials are equal to each other then a particle subject to the potential $\varphi$ should not be accelerating with respect to its proper accelerating reference frame. This is the basic idea behind the method of formulating the dynamical conditions for a proper reference frame of an accelerated observer that we present below.

We can see that the line element Eq.~(\ref{eq:inert-metr-01}) is valid only at the Newtonian level. To extend this metric to the post-Newtonian level, first we write the acceleration as a sum of a Newtonian and post-Newtonian terms:
\begin{equation}
b^\alpha(y^0)=b^{[0]\alpha}(y^0)+c^{-2}b^{[2]\alpha}(y^0)+{\cal O}(c^{-4}).
\label{eq:inert-accel}
\end{equation}
Furthermore, we impose the isotropic harmonic gauge condition on the resulting metric:
{}
\begin{eqnarray}
2c\partial_0 \eta^{[2]\rm eff}_{00} +\partial^\nu \eta^{[3]\rm eff}_{0\nu} &=&
{\cal O}(c^{-2}), \hskip 20pt
 \label{eq:(DeDgGa)in-01+}\\
\eta^{[2]\rm eff}_{\alpha\beta}+\gamma_{\alpha\beta}\eta_{00}^{[2]\rm eff} &=& {\cal O}(c^{-2}).
 \label{eq:(DeDgGa)in-02+}
\end{eqnarray}
This metric represents the combined contributions of the external force (introduced via $b^\alpha$) and the inertial or frame-reaction force (introduced via $u$ and $u^\alpha$) that affect the motion of our observer in the accelerating frame.

With the help of these equations we can reconstruct the general structure of the metric that corresponds to the acceleration (\ref{eq:inert-accel}).
It follows from Eq.~(\ref{eq:(DeDgGa)in-01+}) that the acceleration-induced contribution to the mixed-index components of the harmonic and non-rotating metric $\eta^{[3]\rm eff}_{0\nu}$ may be represented as
\begin{equation}
\eta^{[3]\rm eff}_{0\alpha}=-4\gamma_{\alpha\lambda}\Big(u^\lambda(y)-{\textstyle\frac{1}{3}}\dot\varphi_0 y^\lambda+{\cal O}(y^2)\Big)+{\cal O}(c^{-2}),
\label{eq:totcond1}
\end{equation}
where $\dot\varphi_0=c\partial_0\varphi_0$. As we will be interested in the values of these potentials and their first spatial derivatives on the observer's world-line, we do not need an explicit form of the terms denoted by ${\cal O}(y^2)$.
Furthermore, Eq.~(\ref{eq:(DeDgGa)in-02+}) suggests that the spatial component of the metric tensor  $\eta^{[2]\rm eff}_{\alpha\beta}$ in harmonic coordinates has the form:
\begin{equation}
\eta^{[2]\rm eff}_{\alpha\beta}=2\gamma_{\alpha\beta}\Big(u(y)+\varphi(y)\Big)+{\cal O}(c^{-2}).
\label{eq:totcond2}
\end{equation}

The conditions (\ref{eq:totcond1})--(\ref{eq:totcond2}) allow us to extend the metric (\ref{eq:inert-metr-01}) beyond the Newtonian level. In the fictitious metric $\eta^{\rm eff}_{mn}$ that is introduced to represent the combined effects of an external force and an accelerating frame, we must combine the contributions of the external and the frame-reaction forces:
\begin{eqnarray}
\eta^{\rm eff}_{00}(y)&=& 1-\frac{2}{c^2}\Big(u(y)+\varphi(y)\Big)+\frac{2}{c^4}\Big(\big(u(y)+\varphi(y)\big)^2+{\cal O}(y^2)\Big)+{\cal O}(c^{-6}),
\label{eq:eta_00-cov-pot-f-comb}\\
\eta^{\rm eff}_{0\alpha}(y)&=& -\gamma_{\alpha\lambda}\frac{4}{c^3}
\Big(u^\lambda(y)-{\textstyle\frac{1}{3}}\dot\varphi_0 y^\lambda+{\cal O}(y^2)\Big)+{\cal O}(c^{-5}),
\label{eq:eta_0a-cov-pot-f-comb}\\
\eta^{\rm eff}_{\alpha\beta}(y)&=& \gamma_{\alpha\beta}+\gamma_{\alpha\beta}\frac{2}{c^2} \Big(u(y)+\varphi(y)\Big)+{\cal O}(c^{-4}).
\label{eq:eta_ab-cov-pot-f-comb}
\end{eqnarray}
This metric represents accurately, at the post-Galilean level, the effects of an externally induced acceleration $b^\alpha$ on an observer in an accelerating frame characterized by $u$ and $u^\alpha$, while also satisfying the harmonic gauge.

\subsection{Equation of motion of an accelerating observer}
\label{sec:fict-metr-eq-mot}

The metric tensor $\eta^{\rm eff}_{mn}$ given by Eqs.~(\ref{eq:eta_00-cov-pot-f-comb})--(\ref{eq:eta_ab-cov-pot-f-comb}) allows us to study the dynamics of an observer or test particle that moves in response to the presence of the external force. The test particle Lagrangian $L_{\rm eff}$ that corresponds to this system can be obtained directly from the metric $\eta^{\rm eff}_{mn}$ and written as \cite{Chandra-Contopulos-65}:
{}
\begin{eqnarray}
L_{\rm eff}&=&-mc^2\frac{ds}{dy^0}=-mc^2\Big({\eta^{\rm eff}_{mn}\frac{dy^m}{dy^0}\frac{dy^n}{dy^0}}\Big)^{1/2}=\nonumber\\
&=&-mc^2\Big\{1+c^{-2}\Big({\textstyle\frac{1}{2}}v{}_\epsilon v^\epsilon-\tilde u\Big)+c^{-4}\Big({\textstyle\frac{1}{2}}{\tilde u}^2-{\textstyle\frac{1}{8}}(v{}_\epsilon v^\epsilon)^2-4v_\epsilon{\tilde u}^\epsilon+{\textstyle\frac{3}{2}}v{}_\epsilon v^\epsilon \tilde u+{\cal O}(y^2)\Big)+{\cal O}(c^{-6})\Big\}, \label{eq:lagr}
\end{eqnarray}
where $\tilde u=u(y)+\varphi(y)+{\cal O}(c^{-4})$ and $\tilde u^\alpha=u^\alpha(y)-{\textstyle\frac{1}{3}}\dot\varphi_0 y^\alpha+{\cal O}(c^{-2})$ are the combined scalar and vector potentials that consist of the inertial potentials given by Eqs.~(\ref{eq:pot_loc-w_0-cov}), (\ref{eq:pot_loc-w_a-cov}) representing the inertial frame-reaction potential, and the potentials Eq.~(\ref{eq:scal-pot}) combined with Eq.~(\ref{eq:inert-accel}), representing the external force.

The Lagrangian (\ref{eq:lagr}) leads to the following equation of motion\footnote{It can be shown that Eq.~(\ref{eq:eq-mot}) is equivalent to the geodesic equation written with respect to the metric $\eta^{\rm eff}_{mn}$ for the combined system of external and frame-reaction forces. However, Eq.~(\ref{eq:eq-mot}) has the advantage that it allows to separate relativistic quantities and to study the motion of the system in a more straightforward way. Introducing this equation through the Lagrangian (\ref{eq:lagr}) offers us the opportunity to identify unambiguously dynamical quantities, most notably the canonical and ordinary momenta. Indeed we note that the canonical momentum is given by
$p_{\rm can}^\alpha=\partial L_{\rm eff}/\partial v^\alpha=v^\alpha\left(1+c^{-2}\big(3\tilde u-{\textstyle\frac{1}{2}}v{}_\epsilon v^\epsilon\big)+{\cal O}(c^{-4})\right)-4c^{-2}\tilde u^\alpha.\label{eq:canmon}
$
In direct analogy with electromagnetism \cite{Landau-Lifshitz-1988}, we note that the last term in this equation is an inerto-magnetic term; in contrast, the first group of terms corresponds to the test particle's ordinary (mechanical) momentum:
$
p^\alpha=v^\alpha\left(1+c^{-2}\big(3\tilde u-{\textstyle\frac{1}{2}}v{}_\epsilon v^\epsilon\big)+{\cal O}(c^{-4})\right).\label{eq:ordmon}
$}:
{}
\begin{eqnarray}
c\frac{d}{dy^0}\Big\{v^\alpha\Big(1+c^{-2}\big(3\tilde u
-{\textstyle\frac{1}{2}}v{}_\epsilon v^\epsilon\big)+{\cal O}(c^{-4})\Big)\Big\}&=&\nonumber\\
=-\partial ^\alpha \tilde u
\Big\{1-c^{-2}\Big({\textstyle\frac{3}{2}}v{}_\epsilon v^\epsilon+
\tilde u
\Big)\Big\}&+&
\frac{4}{c^2}c\partial_0 {\tilde u}^\alpha+
\frac{4}{c^2}v_\epsilon\Big(\partial^\epsilon \tilde u^\alpha-
\partial^\alpha \tilde u^\epsilon\Big)+\frac{1}{c^2}{\cal O}(y)+{\cal O}(c^{-4}).
 \label{eq:eq-mot}
\end{eqnarray}

The condition that the test particle is to remain at rest with respect to the accelerating frame, then, amounts to demanding that its ordinary linear momentum be conserved, i.e., that its total time derivative is to remain zero. In other words, we require that there will be no forces acting on the observer in its proper reference frame or the right-hand side of (\ref{eq:eq-mot}) vanishes on its world-line. We assume that the observer is located at the spatial origin, $y^\alpha=0$. This leads to the following equation, constructed from the right-hand side of (\ref{eq:eq-mot}) that is valid on the observer's world-line:
\begin{equation}
\Big\{-\partial ^\alpha \tilde u
\Big\{1-c^{-2}\Big({\textstyle\frac{3}{2}}v{}_\epsilon v^\epsilon+
\tilde u
\Big)\Big\}+
\frac{4}{c^2}c\partial_0 {\tilde u}^\alpha+
\frac{4}{c^2}v_\epsilon\Big(\partial^\epsilon \tilde u^\alpha-
\partial^\alpha \tilde u^\epsilon\Big)+\frac{1}{c^2}{\cal O}(y)+{\cal O}(c^{-4})\Big\}\Big|_{|{\vec y}|\rightarrow0}=0.
\end{equation}
We choose the coordinate system $\{y^m\}$ such that along the observer's world-line the potentials $\tilde u$ and $\tilde u^\alpha$ and their first spatial derivatives vanish, so that the metric $\eta^{\rm eff}_{mn}$ reduces to the Minkowski metric along the world-line. Therefore, we require that the following relations involving the frame-reaction potentials $u$ and $u^\alpha$ hold along the world-line:
{}
\begin{eqnarray}
 \lim_{|{\vec y}|\rightarrow 0} u(y)     &=&-\varphi_0+
{\cal O}(c^{-4}), \qquad~~~\,
\lim_{|{\vec y}|\rightarrow 0} \partial^{}_\beta u(y) = b_\beta+
{\cal O}(c^{-4}),
 \label{eq:fermi-cov_pot-0}\\
 \lim_{|{\vec y}|\rightarrow 0} u^\alpha(y)     &=&
{\cal O}(c^{-2}),  \qquad\qquad\qquad
\lim_{|{\vec y}|\rightarrow 0} \partial^{}_\beta u^\alpha(y)     =
{\textstyle\frac{1}{3}}\delta^\alpha_\beta\dot\varphi_0+{\cal O}(c^{-2}).
 \label{eq:fermi-cov_pot-a}
\end{eqnarray}
As we shall see in the next section, these conditions yield the equations needed to fix the time-like coordinate on the observer's world-line and to determine the explicit form of the coordinate transformation functions ${\cal K}, {\cal L}$ and ${\cal Q}^\alpha$.

\subsection{Application of the dynamical conditions}

Imposing the conditions (\ref{eq:fermi-cov_pot-0})--(\ref{eq:fermi-cov_pot-a}) on the potentials $u$ and $u^\alpha$, which are given by Eqs.~(\ref{eq:pot_loc-w_0-cov})--(\ref{eq:pot_loc-w_a-cov}), results in the following set of partial differential equations set on the observer's world-line:
{}
\begin{eqnarray}
u|^{}_{y=0}+\varphi_0&=&\varphi_0-\frac{\partial \kappa_0}{\partial y^0}-
{\textstyle\frac{1}{2}}v_{0}{}_\epsilon v_{0}^\epsilon -
\frac{1}{c^2}\Big\{ \frac{\partial {\cal L}}{\partial y^0}+cv_{0\epsilon}\frac{\partial {\cal Q}^\epsilon}{\partial y^0}+
{\textstyle\frac{1}{2}}\Big(\frac{\partial \kappa_0}{\partial y^0}\Big)^2-\Big(\frac{\partial \kappa_0}{\partial y^0}+
{\textstyle\frac{1}{2}}v_{0}{}_\epsilon v_{0}^\epsilon\Big)^2\Big\}=
{\cal O}(c^{-4}),\hskip16pt
\label{eq:match-w_0-cov}\\
\partial^{}_\beta u|^{}_{y=0} - b_\beta&=&a^{}_{0\beta} -b^{[0]}_\beta-
\frac{1}{c^2}\Big\{b^{[2]}_\beta+ \frac{\partial^2 {\cal L}}{\partial y^\beta\partial y^0}+cv_{0\epsilon}\frac{\partial^2 {\cal Q}^\epsilon}{\partial y^\beta\partial y^0}+
a_{0\beta}\Big(\frac{\partial \kappa_0}{\partial y^0}+
v_{0}{}_\epsilon v_{0}^\epsilon\Big)\Big\}={\cal O}(c^{-4}),
\hskip24pt
\label{eq:match-w_0dir-cov}\\
u^\alpha|^{}_{y=0}&=&-
{\textstyle\frac{1}{4}}\big\{ \gamma^{\alpha\epsilon}\frac{1}{c}\frac{\partial {\cal L}}{\partial y^\epsilon}+c\frac{\partial {\cal Q}^\alpha}{\partial y^0}+\gamma^{\alpha\epsilon}v_{0\lambda}\frac{\partial {\cal Q}^\lambda}{\partial y^\epsilon}-v_{0}^\alpha\frac{\partial \kappa_0}{\partial y^0}\big\}={\cal O}(c^{-2}),
\label{eq:match-w-cov}\\
\partial^{}_\beta u^{}_\alpha|^{}_{y=0}-{\textstyle\frac{1}{3}}\gamma_{\alpha\beta}\dot\varphi_0&=&
-{\textstyle\frac{1}{3}}\gamma_{\alpha\beta}\dot\varphi_0-
{\textstyle\frac{1}{4}}\big\{ \frac{1}{c}\frac{\partial^2 {\cal L}}{\partial y^\alpha\partial y^\beta}+
c\gamma_{\alpha\lambda}\frac{\partial^2 {\cal Q}^\lambda}{\partial y^0\partial y^\beta}+v_{0\lambda}\frac{\partial^2 {\cal Q}^\lambda}{\partial y^\alpha\partial y^\beta}+v_{0\alpha} a_{0\beta}\big\}={\cal O}(c^{-2}).
\label{eq:match-wdir-cov}
\end{eqnarray}

Eqs.~(\ref{eq:match-w_0-cov})--(\ref{eq:match-wdir-cov}) may be used to determine uniquely the form of the coordinate transformation functions ${\cal K}, {\cal L}$ and ${\cal Q}^\alpha$. Indeed, from the first two equations above, (\ref{eq:match-w_0-cov}) and (\ref{eq:match-w_0dir-cov}), we immediately have:
{}
\begin{eqnarray}
\varphi_0-\frac{\partial \kappa_0}{\partial y^0}- {\textstyle\frac{1}{2}}v_{0}{}_\epsilon v_{0}^\epsilon &=& {\cal O}(c^{-4}), \label{eq:dir0-K}\\
a_{0}^\alpha&=&b^{[0]\alpha} +{\cal O}(c^{-4}).
\label{eq:a-Newton}
\end{eqnarray}

Using Eq.~(\ref{eq:dir0-K}) in Eq.~(\ref{eq:Ka-cov}), we can determine uniquely the function  ${\cal K}$:
{}
\begin{equation}
{\cal K}(y)= \int_{y^0_{0}}^{y^0}\!\!\!
\Big(\varphi_0-{\textstyle\frac{1}{2}}v^{}_{0\epsilon} v_{0}^\epsilon\Big)dy'^0 -
c(v^{}_{0\epsilon} y^\epsilon) + {\cal O}(c^{-4}).
\label{eq:Ka-sol-cov}
\end{equation}

Substituting this expression into Eq. (\ref{eq:d-Q}) we can determine the function ${\cal Q}^\alpha$:
\begin{eqnarray}
{\cal Q}^\alpha(y)&=& q^{\alpha}_{0} -
\Big({\textstyle\frac{1}{2}}v^\alpha_{0} v^\epsilon_{0}+
\omega_{0}^{\alpha\epsilon}+\gamma^{\alpha\epsilon} \varphi_0\Big){y}_\epsilon +
{a_{0}}_\epsilon\Big(y^\alpha y^\epsilon-
{\textstyle\frac{1}{2}}\gamma^{\alpha\epsilon}{y}_\lambda y^\lambda\Big) + {\cal O}(c^{-2}).
\label{eq:d-Q=}
\end{eqnarray}

Finally, the general solution for the function ${\cal L}$ given by Eq.~(\ref{eq:L-gen-cov}) now takes the following form:
{}
\begin{eqnarray}
{\cal L}(y)&=&\ell_{0}+\ell_{0\lambda}\,y^\lambda+
{\textstyle\frac{1}{2}}\ell_{0\lambda\mu}\,y^\lambda y^\mu
+{\textstyle\frac{1}{6}}c\Big(
\big(v^{}_{0\epsilon} a_{0}^\epsilon\big)-\dot\varphi_0\Big)(y_\nu y^\nu)+
{\textstyle\frac{1}{10}}c(\dot{a}_{0\epsilon}y^\epsilon)
(y_\nu y^\nu)+\delta\ell(y)+{\cal O}(c^{-2}).
\label{eq:L-gen-cov+}
\end{eqnarray}

The next task is to find the remaining undetermined time-dependent functions present in ${\cal Q}^\alpha$ and ${\cal L}$, as given by Eqs.~(\ref{eq:d-Q=}) and (\ref{eq:L-gen-cov+}). To do this, we rewrite the remaining parts of Eqs.~(\ref{eq:match-w_0-cov})--(\ref{eq:match-wdir-cov}) as a system of partial differential equations with respect to ${\cal L}$, again set on the observer's world-line:
{}
\begin{eqnarray}
\frac{\partial {\cal L}}{\partial y^0}+cv_{0\epsilon}\frac{\partial {\cal Q}^\epsilon}{\partial y^0}+
{\textstyle\frac{1}{2}}\Big(\frac{\partial \kappa_0}{\partial y^0}\Big)^2-\Big(\frac{\partial \kappa_0}{\partial y^0}+
{\textstyle\frac{1}{2}}v_{0}{}_\epsilon v_{0}^\epsilon\Big)^2 &=&
{\cal O}(c^{-2}),\hskip24pt
\label{eq:match-w_0-cov-L}\\
\frac{\partial^2 {\cal L}}{\partial y^\beta\partial y^0}+cv_{0\epsilon}\frac{\partial^2 {\cal Q}^\epsilon}{\partial y^\beta\partial y^0}+
a_{0\beta}\Big(\frac{\partial \kappa_0}{\partial y^0}+
v_{0}{}_\epsilon v_{0}^\epsilon\Big)
&=&-b^{[2]}_\beta+{\cal O}(c^{-2}),\hskip24pt
\label{eq:match-w_0dir-cov-L}\\
\gamma^{\alpha\epsilon}\frac{1}{c}\frac{\partial {\cal L}}{\partial y^\epsilon}+\gamma^{\alpha\epsilon}v_{0\lambda}\frac{\partial {\cal Q}^\lambda}{\partial y^\epsilon}+c\frac{\partial {\cal Q}^\alpha}{\partial y^0}-v_{0}^\alpha\frac{\partial \kappa_0}{\partial y^0}&=&{\cal O}(c^{-2}),
\label{eq:match-w-cov-L}\\
\frac{1}{c}\frac{\partial^2 {\cal L}}{\partial y^\alpha\partial y^\beta}+v_{0\lambda}\frac{\partial^2 {\cal Q}^\lambda}{\partial y^\alpha\partial y^\beta}+ c\gamma_{\alpha\lambda}\frac{\partial^2 {\cal Q}^\lambda}{\partial y^0\partial y^\beta}+v_{0\alpha} a_{0\beta}&=&-{\textstyle\frac{4}{3}}\gamma_{\alpha\beta}\dot\varphi_0+{\cal O}(c^{-2}).
\label{eq:match-wdir-cov-L}
\end{eqnarray}
The equations above are may be used to determine the remaining unknown time-dependent functions $\ell_{0}$, $\ell_{0\lambda}$,  $\ell_{0\lambda\mu}$, and also $q^{\alpha}_{0}$ and $\omega_{0}^{\alpha\epsilon}$ still present in the coordinate transformation functions. Thus, substituting the previously obtained solutions for ${\cal K}$ and ${\cal Q}^\alpha$, given by Eqs.~(\ref{eq:Ka-sol-cov}) and (\ref{eq:d-Q=}), in Eq.~(\ref{eq:match-w_0-cov-L}) leads to the following solution for ${\dot \ell}_{0}$:
{}
\begin{eqnarray}
{\textstyle\frac{1}{c}}\dot\ell_{0}&=&
-v_{0\epsilon}{\dot q}^\epsilon_{0}-
{\textstyle\frac{1}{8}}(v_{0\epsilon}v^\epsilon_{0})^2+
{\textstyle\frac{1}{2}}(v_{0\epsilon}v^\epsilon_{0})\varphi_0+
{\textstyle\frac{1}{2}}\varphi_0^2+{\cal O}(c^{-2}).
\label{eq:ell_0}
\end{eqnarray}

Next, Eq.~(\ref{eq:match-w_0dir-cov-L}) results in the equation for $\dot\ell^\alpha_{0}$:
{}
\begin{eqnarray}
{\textstyle\frac{1}{c}}\dot\ell^\alpha_{0}&=&-b^{[2]\alpha}+
{\textstyle\frac{1}{2}}v_0^\alpha\big(v_{0\epsilon}a^\epsilon_{0}\big)
-v^{}_{0\epsilon}{\dot\omega}_{0}^{\alpha\epsilon}+
v^\alpha_0\dot\varphi_0-a^\alpha_0\varphi_0+{\cal O}(c^{-2}).
\label{eq:ell-dot}
\end{eqnarray}

From Eq.~(\ref{eq:match-w-cov-L}) we can determine $\ell^\alpha_{0}$:
{}
\begin{eqnarray}
{\textstyle\frac{1}{c}}\ell^\alpha_{0}&=&-{\dot q}^\alpha_{0}-
v^{}_{0\epsilon}\,\omega_{0}^{\alpha\epsilon}+2v^\alpha_0\varphi_0+{\cal O}(c^{-2}).
\label{eq:ell}
\end{eqnarray}

Eq.~(\ref{eq:match-wdir-cov-L}) leads to the following solution for $\ell^{\alpha\beta}_{0}$:
{}
\begin{eqnarray}
{\textstyle\frac{1}{c}}\ell^{\alpha\beta}_{0}&=&
-{\textstyle\frac{3}{2}}v^\alpha_{0}a^\beta_{0}-
{\textstyle\frac{1}{2}}v^\beta_{0}a^\alpha_{0}+
{\textstyle\frac{2}{3}}\gamma^{\alpha\beta}(a^{}_{0\epsilon}v^\epsilon_{0})+{\dot\omega}_{0}^{\alpha\beta}+
{\cal O}(c^{-2}).
\label{eq:ellb}
\end{eqnarray}
The quantity $\ell^{\alpha\beta}_{0}$ is an STF tensor. The expression on the right-hand side must, therefore, be also symmetric. This can be achieved by choosing the anti-symmetric tensor ${\dot\omega}_{0}^{\alpha\beta}$ appropriately. This can be done uniquely, resulting in
\begin{equation}
\dot\omega_{0}^{\alpha\beta}=
{\textstyle\frac{1}{2}}\big(v^{\alpha}_{0}a^{\beta}_{0}-v^{\beta}_{0}a^{\alpha}_{0}\big)+{\cal O}(c^{-2}),
\label{eq:omega}
\end{equation}
which is the relativistic Thomas precession \cite{Brumberg-book-1991}. Using this expression yields the following solution for $\ell^{\alpha\lambda}_{0}$:
{}
\begin{eqnarray}
{\textstyle\frac{1}{c}}\ell^{\alpha\beta}_{0}&=&
-v^\alpha_{0}a^\beta_{0}-v^\beta_{0}a^\alpha_{0}+
{\textstyle\frac{2}{3}}\gamma^{\alpha\beta}v_{0\epsilon}a_{0}^\epsilon
+{\cal O}(c^{-2}).
\label{eq:ellb-dd}
\end{eqnarray}
Furthermore, we can use Eq.~(\ref{eq:omega}) in Eq.~(\ref{eq:ell-dot}), leading to the solution for $\dot\ell^\alpha_{0}$ in the form:
{}
\begin{eqnarray}
{\textstyle\frac{1}{c}}\dot\ell^\alpha_{0}&=&-b^{[2]\alpha}+
{\textstyle\frac{1}{2}}a_0^\alpha\big(v^{}_{0\epsilon}v^\epsilon_{0}\big)+v^\alpha_0\dot\varphi_0-a^\alpha_0\varphi_0
+{\cal O}(c^{-2}).
\label{eq:ell-dot-2}
\end{eqnarray}

Eqs.~(\ref{eq:ell}) and (\ref{eq:ell-dot-2}) allow us to determine $q_{0}^{\alpha}$. Specifically, rewriting Eq.~(\ref{eq:ell-dot-2}) as
{}
\begin{eqnarray}
{\textstyle\frac{1}{c}}\dot\ell^\alpha_{0}&=&-b^{[2]\alpha}+
{\textstyle\frac{1}{2}}\Big(v_0^\alpha\big(v^{}_{0\epsilon}v^\epsilon_{0}\big)\Big)^{\!.}-v_0^\alpha\big(v^{}_{0\epsilon}a^\epsilon_{0}\big)+\Big(v_0^\alpha\varphi_0\Big)^{\!.}-2a^\alpha_0\varphi_0
+{\cal O}(c^{-2})
\label{eq:ell-dot-2a}
\end{eqnarray}
and formally integrating it with respect to $y^0$, we obtain another expression for $\ell^\alpha_{0}$:
{}
\begin{eqnarray}
{\textstyle\frac{1}{c}}\ell^\alpha_{0}&=&
{\textstyle\frac{1}{2}}v_0^\alpha\big(v^{}_{0\epsilon}v^\epsilon_{0}\big)+v_0^\alpha\varphi_0-\int \Big(b^{[2]\alpha}+v_0^\alpha\big(v^{}_{0\epsilon}a^\epsilon_{0}\big)+2a^\alpha_0\varphi_0\Big){\textstyle\frac{1}{c}}dy^0
+{\cal O}(c^{-2}).
\label{eq:ell-dot-2b}
\end{eqnarray}
Eqs.~(\ref{eq:ell}) and (\ref{eq:ell-dot-2b}) can now be solved with respect to  ${\dot q}^\alpha_{0}$:
{}
\begin{eqnarray}
\dot q_{0}^{\alpha}&=&  -{\textstyle\frac{1}{2}}v_0^\alpha\big(v^{}_{0\epsilon}v^\epsilon_{0}\big)-v^{}_{0\epsilon}\,\omega_{0}^{\alpha\epsilon}+v_0^\alpha\varphi_0+\int \Big(b^{[2]\alpha}+v_0^\alpha\big(v^{}_{0\epsilon}b^{[0]\epsilon}\big)+2b^{[0]\alpha}\varphi_0\Big){\textstyle\frac{1}{c}}dy^0
+{\cal O}(c^{-2}),
\label{eq:q-dot}
\end{eqnarray}
where we used Eq.~(\ref{eq:a-Newton}) for $a_0^\alpha$. The first term in this expression is the Lorentzian factor, while the other terms explicitly depend on the observer's acceleration and the value of the external potential $\varphi_0=\varphi_0(y^0)$ on its world-line.

Finally, differentiating Eq.~(\ref{eq:ell}) with respect to time and subtracting the result from Eq.~(\ref{eq:ell-dot-2}) (or just simply differentiating Eq.~(\ref{eq:q-dot}) with respect to time), we obtain the following equation on $\ddot q_{0}^{\alpha}$:
{}

\begin{eqnarray}
\ddot q_{0}^{\alpha}&=&  b^{[2]\alpha} -\Big(
{\textstyle\frac{1}{2}}v^\alpha_{0}v_0^\epsilon +\omega_{0}^{\alpha\epsilon}\Big)a_{0\epsilon}+3a^\alpha_0\varphi_0+v^\alpha_0\dot\varphi_0+
{\cal O}(c^{-2}).
\label{eq:q-ddot}
\end{eqnarray}
Therefore, with the knowledge of the external acceleration $b^{\alpha}$ and background potential $\varphi_0$, we can use Eqs.~(\ref{eq:q-dot}) and (\ref{eq:q-ddot}) to completely determine the function $q_{0}^{\alpha}$.

The true position of a test particle includes terms to all orders, not just the first-order (Galilean) term. This led us to introduce the vector, $x^\alpha_0(y^0)$, defined by Eq.~(\ref{eq:totvec}). Combining this definition with Eqs.~(\ref{eq:a-Newton}) and (\ref{eq:q-ddot}), we can now write the magnitude of the frame-reaction force (acting on the unit mass) written in the local coordinates of the accelerated observer as it relates to the measured acceleration $b^\alpha$:
\begin{eqnarray}
\ddot x^\alpha_{0}(y^0)=  a^\alpha_{0}+c^{-2}\ddot q_{0}^{\alpha} +
{\cal O}(c^{-4})&=&
b^{[0]\alpha}+\frac{1}{c^2}\Big\{b^{[2]\alpha}-\Big(
{\textstyle\frac{1}{2}}v^\alpha_{0}v_0^\epsilon +\omega_{0}^{\alpha\epsilon}\Big)b^{[0]}_\epsilon
+3a^\alpha_0\varphi_0+v^\alpha_0\dot\varphi_0\Big\}+
{\cal O}(c^{-4})\nonumber\\
&=& b_\epsilon\Big\{\gamma^{\alpha\epsilon}-\frac{1}{c^2}\Big(
{\textstyle\frac{1}{2}}v^\alpha_{0}v_0^\epsilon +\omega_{0}^{\alpha\epsilon}\Big)
\Big\}+
\frac{1}{c^2}\Big\{3a^\alpha_0\varphi_0+v^\alpha_0\dot\varphi_0\Big\}+
{\cal O}(c^{-4}).
\label{eq:geod_eq-local-cov}
\end{eqnarray}

The equation of motion (\ref{eq:geod_eq-local-cov}) establishes the correspondence between $b^\alpha$, the externally-induced acceleration of the observer, and the fictitious frame-reaction acceleration $\ddot x^\alpha_0(y^0)$ that is needed to keep the observer at rest in its proper reference frame. This frame-reaction force balances the effect of the external inertial force acting on the observer.

\subsection{Summary of results for the direct transformations}
\label{sec:sumdirect}

We sought general post-Galilean transformations between the Minkowski frame $\{x^k\}$ and the dynamically non-rotating coordinates $\{y^k\}$ of a proper reference frame of an accelerated observer. We did that representing such a coordinate transformation in the most general form:
\begin{eqnarray}
x^0&=& y^0+c^{-2}{\cal K}(y^0,y^\epsilon)+c^{-4}{\cal L}(y^0,y^\epsilon)+O(c^{-6}),\\[3pt]
x^\alpha&=& y^\alpha+z^\alpha_{0}(y^0)+
c^{-2}{\cal Q}^\alpha(y^0,y^\epsilon)+O(c^{-4}).
\end{eqnarray}

To determine the unknown functions ${\cal K}$, ${\cal L}$, and ${\cal Q}^\alpha$, we used the following approach:
\begin{inparaenum}[i)]
\item we imposed the harmonic gauge conditions on the accelerated  Minkowski metric, the metric tensor in the local accelerating frame;
\item we ensured that the accelerating frame is non-rotating and the chosen coordinates are spatially isotropic;
\item we introduced an accelerating observer at rest with respect to its proper accelerating frame;
\item we cast the combination of all forces acting on the accelerating observer in the form of a fictitious metric that includes the external and frame-reaction potentials;
\item we imposed the harmonic gauge on the fictitious metric;
\item we required
that a co-moving test particle's ordinary three-dimensional linear momentum be conserved on the world-line occupied by the accelerating frame.
\end{inparaenum}

Together, these conditions were sufficient to determine ${\cal K}$, ${\cal L}$ and ${\cal Q}^\alpha$ unambiguously:
{}
\begin{eqnarray}
{\cal K}(y)&=& \!\!\int_{y^0_{0}}^{y^0}\!\!\!
\Big(\varphi_0-{\textstyle\frac{1}{2}}v^{}_{0\epsilon} v_{0}^\epsilon\Big)dy'^0 -
c(v^{}_{0\epsilon} y^\epsilon)+{\cal O}(c^{-4}),
\label{eq:fin-sol-K}\\
{}
{\cal L}(y)&=& -\!\!\int_{y^0_{0}}^{y^0}\!\!\!
\Big(v_{0\epsilon}{\dot q}^\epsilon_{0}+{\textstyle\frac{1}{8}}(v_{0\epsilon}v^\epsilon_{0})^2-
{\textstyle\frac{1}{2}}(v_{0\epsilon}v^\epsilon_{0})\varphi_0-{\textstyle\frac{1}{2}}\varphi_0^2\Big)dy'^0
-c\Big({\dot q}_{0\epsilon}+
v_0^\lambda\,\omega_{0\epsilon\lambda}-2v_{0\epsilon}\varphi_0\Big)y^\epsilon
-\nonumber\\
&&-{\textstyle\frac{1}{2}}c\Big(a^{}_{0\epsilon}v^{}_{0\lambda}+
a^{}_{0\lambda}v^{}_{0\epsilon}-\gamma_{\epsilon\lambda}
a_{0\mu}v_0^\mu+{\textstyle\frac{1}{3}}\gamma_{\epsilon\lambda}\dot\varphi_0\Big)y^\epsilon y^\lambda+{\textstyle\frac{1}{10}}c(\dot{a}_{0\epsilon}y^\epsilon)(y_\nu y^\nu)
+\delta\ell(y)+{\cal O}(c^{-2}),\label{eq:Lfinal}\\
{\cal Q}^\alpha(y)&=& q^{\alpha}_{0} -
\Big({\textstyle\frac{1}{2}}v^\alpha_{0} v^\epsilon_{0}+
\omega_{0}^{\alpha\epsilon}+\gamma^{\alpha\epsilon}\varphi_0\Big){y}_\epsilon +{a_{0}}_\epsilon\Big(y^\alpha y^\epsilon-
{\textstyle\frac{1}{2}}\gamma^{\alpha\epsilon}{y}_\lambda y^\lambda\Big)+{\cal O}(c^{-2}),
\end{eqnarray}
with the anti-symmetric relativistic precession matrix $\omega_{0}^{\alpha\lambda}$ given by Eq.~(\ref{eq:omega}), and the post-Newtonian component of the spatial coordinate in the local frame, $q_{0}^{\alpha}$, given by Eq.~(\ref{eq:q-ddot}).

Substituting these solutions for the functions ${\cal K}$, ${\cal L}$, and ${\cal Q}^\alpha$ into the expressions for the inertial frame-reaction potentials $u$ and $u^\alpha$ given by Eqs.~(\ref{eq:pot_loc-w_0-cov})--(\ref{eq:pot_loc-w_a-cov}), we find the following form for these potentials:
{}
\begin{eqnarray}
u(y)&=&(a_\epsilon y^\epsilon)-\varphi_0+\frac{1}{c^2}\Big\{
{\textstyle\frac{1}{2}}\Big(3y^\epsilon y^\lambda
-\gamma^{\epsilon\lambda} y_\mu y^\mu\Big)
a^{}_{0\epsilon} a^{}_{0\lambda} -
{\textstyle\frac{1}{10}}
(\ddot a^{}_{0\epsilon} y^\epsilon)(y_\mu y^\mu)
+{\textstyle\frac{1}{6}}\ddot\varphi_0(y_\mu y^\mu)
-\partial_0 \delta\ell\Big\}+{\cal O}(c^{-4}),~~~
\label{eq:pot_loc-w_0-cov+}\\[3pt]
u^\alpha(y)&=&-
{\textstyle\frac{1}{10}}\big(3y^\alpha y^\epsilon-
\gamma^{\alpha\epsilon}y_\mu y^\mu\big){\dot a}^{}_{0\epsilon}
+{\textstyle\frac{1}{3}}\dot\varphi_0 y^\alpha-\partial^\alpha {\textstyle\frac{1}{4c}}\delta\ell+{\cal O}(c^{-2}),
\label{eq:pot_loc-w_a-cov+}
\end{eqnarray}
where $a^\alpha$ denotes the frame-reaction acceleration which is equal to the measured external acceleration $b^\alpha$ given by Eq.~(\ref{eq:inert-accel}) or $a^\alpha(y^0)\equiv b^\alpha(y^0)=b^{[0]\alpha}+c^{-2}b^{[2]\alpha}+{\cal O}(c^{-4})$. Substituting these expressions for the inertial potentials into Eqs.~(\ref{eq:eta_00-cov})--(\ref{eq:eta_ab-cov}) leads to the following form of the accelerated Minkowski metric of the arbitrarily moving observer:
{}
\begin{eqnarray}
\eta_{00}(y)&=& 1-\frac{2}{c^2}\Big\{(a_\epsilon y^\epsilon)-\varphi_0\Big\}+\frac{2}{c^4}\Big\{\big((a^{}_{0\epsilon} y^\epsilon)-\varphi_0\big)^2-
{\textstyle\frac{1}{2}}\Big(3y^\epsilon y^\lambda
-\gamma^{\epsilon\lambda} y_\mu y^\mu\Big)
a^{}_{0\epsilon} a^{}_{0\lambda} - \nonumber\\
&&\hskip 158pt+~
{\textstyle\frac{1}{10}}
(\ddot a^{}_{0\epsilon} y^\epsilon)(y_\mu y^\mu)-
{\textstyle\frac{1}{6}}\ddot\varphi_0(y_\mu y^\mu)+\partial_0 \delta\ell\Big\}+{\cal O}(c^{-6}),
\label{eq:eta_00-cov-fin}\\
\eta_{0\alpha}(y)&=& \gamma_{\alpha\lambda}\frac{4}{c^3}\Big\{
{\textstyle\frac{1}{10}}\big(3y^\lambda y^\epsilon-
\gamma^{\lambda\epsilon}y_\mu y^\mu\big){\dot a}_{0\epsilon}
-{\textstyle\frac{1}{3}}\dot\varphi_0 y^\lambda\Big\}+\frac{1}{c^4}\partial_\alpha \delta\ell+{\cal O}(c^{-5}),
\label{eq:eta_0a-cov-fin}\\
\eta_{\alpha\beta}(y)&=& \gamma_{\alpha\beta}+\gamma_{\alpha\beta}\frac{2}{c^2} \Big\{(a^{}_{0\epsilon} y^\epsilon)-\varphi_0\Big\}+{\cal O}(c^{-4}).
\label{eq:eta_ab-cov-fin}
\end{eqnarray}
All terms in this metric are determined except for the function $\delta\ell$, which remains unknown. We note that the potentials Eqs.~(\ref{eq:pot_loc-w_0-cov+})--(\ref{eq:pot_loc-w_a-cov+}) depend on the partial derivatives of $\delta \ell$. The same partial derivatives appear in the temporal and mixed components of the metric (\ref{eq:eta_00-cov-fin})--(\ref{eq:eta_0a-cov-fin}). The presence of these terms in the metric amounts to adding a full time derivative to the Lagrangian that describes the system of the moving observer. Indeed, separating in the Lagrangian constructed from Eqs.~(\ref{eq:eta_00-cov-fin})--(\ref{eq:eta_ab-cov-fin}) the terms that depend on $\delta\ell$,  we have:
\begin{equation}
\delta L_{\delta\ell}=\frac{2}{c^4}\Big\{\frac{\partial \delta\ell}{\partial {y^0}} +\frac{v^\epsilon}{c}
\frac{\partial \delta\ell}{\partial {y^\epsilon}}\Big\}+{\cal O}(c^{-6})=\frac{2}{c^4}\frac{d\delta\ell}{dy^0}+{\cal O}(c^{-6}).
\end{equation}
As a result, the remainder of the gauge transformation depending on $\delta\ell$ will not change the dynamics in the system and, thus, it can be omitted. After some re-arranging, the frame-reaction potentials $u$ and $u^\alpha$ take the form:
{}
\begin{eqnarray}
u(y)&=&(a_\epsilon y^\epsilon)-\varphi_0+\frac{1}{c^2}\Big\{
{\textstyle\frac{1}{2}}\Big(3a^{}_{0\epsilon}a^{}_{0\lambda}
-\gamma_{\epsilon\lambda} a^{}_{0\mu} a^{\mu}_{0}+
{\textstyle\frac{1}{3}}\gamma_{\epsilon\lambda}\ddot\varphi_0\Big)
y^\epsilon y^\lambda  -
{\textstyle\frac{1}{10}}
(\ddot a^{}_{0\epsilon} y^\epsilon)(y_\mu y^\mu)
\Big\}+{\cal O}(c^{-4}),
\label{eq:pot_loc-w_0-cov+fin}\\[3pt]
u^\alpha(y)&=&-
{\textstyle\frac{1}{10}}\big(3y^\alpha y^\epsilon-
\gamma^{\alpha\epsilon}y_\mu y^\mu\big){\dot a}^{}_{0\epsilon}
+{\textstyle\frac{1}{3}}\dot\varphi_0 y^\alpha
+{\cal O}(c^{-2}).
\label{eq:pot_loc-w_a-cov+fin}
\end{eqnarray}
We can now present the metric of the moving the observer in the following final form:
{}
\begin{eqnarray}
\eta_{00}(y)&=& 1-\frac{2}{c^2}\Big\{(a_\epsilon y^\epsilon)-\varphi_0\Big\}+\frac{2}{c^4}\Big\{\big((a^{}_{0\epsilon} y^\epsilon)-\varphi_0\big)^2-\nonumber\\
&&\hskip 65pt-~
{\textstyle\frac{1}{2}}\Big(3a^{}_{0\epsilon}a^{}_{0\lambda}
-\gamma_{\epsilon\lambda} a^{}_{0\mu} a^{\mu}_{0}+
{\textstyle\frac{1}{3}}\gamma_{\epsilon\lambda}\ddot\varphi_0\Big)
y^\epsilon y^\lambda+
{\textstyle\frac{1}{10}}
(\ddot a^{}_{0\epsilon} y^\epsilon)(y_\mu y^\mu)\Big\}+{\cal O}(c^{-6}),
\label{eq:eta_00-cov-fin+}\\
\eta_{0\alpha}(y)&=& \gamma_{\alpha\lambda}\frac{4}{c^3}\Big\{
{\textstyle\frac{1}{10}}\big(3y^\lambda y^\epsilon-
\gamma^{\lambda\epsilon}y_\mu y^\mu\big){\dot a}_{0\epsilon}
-{\textstyle\frac{1}{3}}\dot\varphi_0 y^\lambda\Big\}+{\cal O}(c^{-5}),
\label{eq:eta_0a-cov-fin+}\\
\eta_{\alpha\beta}(y)&=& \gamma_{\alpha\beta}+\gamma_{\alpha\beta}\frac{2}{c^2} \Big\{(a^{}_{0\epsilon} y^\epsilon)-\varphi_0\Big\}+{\cal O}(c^{-4}).
\label{eq:eta_ab-cov-fin+}
\end{eqnarray}
The coordinate transformations that place the observer into this reference frame are given below:
{}
\begin{eqnarray}
x^0&=& y^0+c^{-2}\Big\{\!\!\int_{y^0_{0}}^{y^0}\!\!\!
\Big(\varphi_0-{\textstyle\frac{1}{2}}(v^{}_{0\epsilon} +c^{-2}{\dot q}^{}_{0\epsilon})(v_{0}^\epsilon+c^{-2}{\dot q}^\epsilon_{0})+c^{-2}\big({\textstyle\frac{1}{2}}\varphi_0^2 +
{\textstyle\frac{1}{2}}(v_{0\epsilon}v^\epsilon_{0})\varphi_0-
{\textstyle\frac{1}{8}}(v_{0\epsilon}v^\epsilon_{0})^2 \big)\Big)dy'^0 +\nonumber\\
&&\hskip 40pt {}-
c\,\Big(v^{\epsilon}_{0} +c^{-2}{\dot q}^{\epsilon}_{0}\Big)\Big(\gamma_{\epsilon\lambda}-
c^{-2}\big(\omega^{}_{0\epsilon\lambda}+2\gamma_{\epsilon\lambda}\varphi_0\big)\Big) y^\lambda\Big\}
-\nonumber\\
&&~~~{}-c^{-4}\Big\{{\textstyle\frac{1}{2}}c\Big(a^{}_{0\epsilon}v^{}_{0\lambda}+ a^{}_{0\lambda}v^{}_{0\epsilon}-\gamma_{\epsilon\lambda}
a_{0\mu}v_0^\mu +{\textstyle\frac{1}{3}}\gamma_{\epsilon\lambda}\dot\varphi_0\Big)y^\epsilon y^\lambda-{\textstyle\frac{1}{10}}c(\dot{a}_{0\epsilon}y^\epsilon)(y_\nu y^\nu)\Big\}+{\cal O}(c^{-6}),
\label{eq:transfrom-fin-0}\\
x^\alpha&=& y^\alpha+z^\alpha_{0}+
c^{-2}\Big\{q^{\alpha}_{0}-
\Big({\textstyle\frac{1}{2}}v^\alpha_{0} v^\epsilon_{0}+
\omega_{0}^{\alpha\epsilon}+\gamma^{\alpha\epsilon}\varphi_0\Big){y}_\epsilon +a^{}_{0\epsilon}\Big(y^\alpha y^\epsilon-
{\textstyle\frac{1}{2}}\gamma^{\alpha\epsilon}{y}_\lambda y^\lambda\Big)\Big\}+{\cal O}(c^{-4}).
\label{eq:transfrom-fin-a}
\end{eqnarray}

The presence of $\varphi_0$ in the metric tensor (\ref{eq:eta_00-cov-fin+})--(\ref{eq:eta_ab-cov-fin+}) and coordinate transformations (\ref{eq:transfrom-fin-0})--(\ref{eq:transfrom-fin-a}) is quite interesting. It shows that in the case of a time-varying background potential $\varphi_0(y^0)$ and no external acceleration, the metric tensor of the corresponding space-time differs from the Minkowski metric. Furthermore, the new space-like coordinates are scaled by $\varphi_0$ and the time-like ones are stretched by both $\varphi_0$ and $\dot\varphi_0$. The result may be intuitive, but was not available previously.
One can verify that in the case of uniform constant velocity motion ($a^\alpha_{0}=0$) and in the absence of the external background potential ($\varphi_0=0$), the metric given by Eqs.~(\ref{eq:eta_00-cov-fin+})--(\ref{eq:eta_ab-cov-fin+}) reduces to the Minkowski metric, $\eta_{mn}=\gamma_{mn}$.  Also, according to Eq.~(\ref{eq:q-dot}) and setting $\varphi_0=0$, the function $\dot q_{0}^{\alpha}$ becomes ${\dot q}_{0}^{\alpha}=  -{\textstyle\frac{1}{2}}v_0^\alpha\big(v^{}_{0\epsilon}v^\epsilon_{0}\big)+{\cal O}(c^{-2})$, and the transformations Eqs.~(\ref{eq:transfrom-fin-0})--(\ref{eq:transfrom-fin-a}) above reduce to the Lorentz transformations.

The expressions (\ref{eq:eta_00-cov-fin+})--(\ref{eq:eta_ab-cov-fin+}) represent the harmonic metric tensor in the local coordinates of the accelerating reference frame. This metric and the transformations (\ref{eq:transfrom-fin-0})--(\ref{eq:transfrom-fin-a}) are new and extend previous formulations obtained with different methods. We were able to derive for the first time an explicit form of the metric tensor corresponding to the space-time of an accelerated observer under harmonic gauge conditions and corresponding coordinate transformations. These results may be verified in laboratory conditions, for instance, those involving high-energy accelerators or precision physical measurements. The formulation can be used to develop models for high precision experiments (for example, those discussed in \cite{Turyshev:2003wt,Turyshev:2007qy,Turyshev:2008ur,Sazhin-etal-2010,Turyshev:2010gk}) where one would needed to relate various observable quantities that are critical for experimental success. However, for a complete description of these experiments we would need to establish the inverse coordinate transformations -- the task that will be performed in the next section.

\section{Inverse transformations}
\label{sec:inverse-tr}

In the preceding sections, we constructed an explicit form of the direct transformation between inertial and accelerating reference frames by applying the harmonic gauge and dynamical conditions on the metric. Given the Jacobian matrix (\ref{eq:(C1a)})--(\ref{eq:(C1d)}), it was most convenient to work with the covariant form of the metric tensor, which could be expressed in terms of the accelerating coordinates by trivial application of the tensor transformation rules (\ref{eq:eta-loc_cov}).

The same logic suggests that if we were to work on the inverse transformation: that is, when it is the inverse Jacobian matrix $\partial y^m/\partial x^n$ that is given in explicit form, it is more convenient to work with the contravariant form of the metric tensor, to which this Jacobian can be applied readily. This simple observation leads us to the idea that we can get the inverse transformations---i.e., from the accelerated to the inertial frame---by simply repeating the previous calculations, but with the contravariant form of the metric tensor instead of the covariant form.

In this section, we show that this is indeed feasible, and accomplish something not usually found in the literature: construction of a method that can be applied for both direct and inverse transformations between inertial and accelerating reference frames at the same time, in a self-consistent manner.

\subsection{General form of the post-Galilean coordinate transformations}

We write the inverse of the general post-Galilean transformations (\ref{eq:trans-0})--(\ref{eq:trans-a}) between the dynamically non-rotating coordinates of accelerating $\{y^m\}$ and those of inertial $\{x^m\}$ reference frames in the following form:
\begin{eqnarray}
y^0&=& x^0+c^{-2}\hat{\cal K}(x^0,x^\epsilon)+c^{-4}\hat{\cal L}(x^0,x^\epsilon)+{\cal O}(c^{-6}),\label{eq:trans-0_inv}\\[3pt]
y^\alpha&=& x^\alpha-z^\alpha_{0}(x^0)+c^{-2}\hat{\cal Q}^\alpha(x^0,x^\epsilon)+{\cal O}(c^{-4}),
\label{eq:trans-a_inv}
\end{eqnarray}
where $z^\mu_{0}(x^0)$ is the Galilean vector connecting the spatial origins of the frames, expressed as a function of global time, $x^0$.
Our objective is to determine the functions $\hat{\cal K}, \hat{\cal L}$ and $\hat{\cal Q}^\alpha$ in explicit form.

We can verify that, in order for Eqs.~(\ref{eq:trans-0_inv})--(\ref{eq:trans-a_inv}) to be inverse to Eqs.~(\ref{eq:trans-0})--(\ref{eq:trans-a}), the ``hatted'' functions ($\hat{\cal K},\hat{\cal L},\hat{\cal Q}^\alpha$) must relate to the original set of (${\cal K},{\cal L},{\cal Q}^\alpha$) via the following expressions:
\begin{eqnarray}
\hat{\cal K}(x)&=& -{\cal K}(x^0,r^\epsilon)+{\cal O}(c^{-4}),
\label{eq:K-hat}\\[3pt]
\hat{\cal L}(x)&=&{\partial_0}{\cal K}(x^0,r^\epsilon) \cdot{\cal K}(x^0,r^\epsilon)+
{\partial_\lambda}{\cal K}(x^0,r^\epsilon) \cdot{\cal Q}^\lambda(x^0,r^\epsilon)-{\cal L}(x^0,r^\epsilon)+{\cal O}(c^{-2}),
\label{eq:L-hat}\\[3pt]
\hat{\cal Q}^\alpha(x)&=&(v^\alpha_0/c)
\,{\cal K}(x^0,r^\epsilon)-{\cal Q}^\alpha(x^0,r^\epsilon)+{\cal O}(c^{-2}),\label{eq:Q-hat}
\end{eqnarray}
with $r^\epsilon=x^\epsilon-x^\epsilon_{0}$, where $x^\mu_{0}(x^0)$ is the post-Galilean vector between the origins of two non-rotating frames expressed as a function of the global time-like coordinate $x^0$ (as opposed to Eq.~(\ref{eq:totvec}), which is given in local time $y^0$) defined as
\begin{equation}
x^\mu_{0}=z^\mu_{0}-c^{-2}\hat{\cal Q}^\mu(x^0,0)+{\cal O}(c^{-4}),
\label{eq:x_0Q}
\end{equation}
and also, $v^\alpha_{0}=\dot z^\alpha_{0}$,  $\partial_0= \partial/\partial x^0$ and $\partial_\lambda= \partial/\partial x^\lambda$, and $x\equiv(x^0,x^\epsilon)$.

The inverse of the Jacobian matrix (\ref{eq:(C1a)})--(\ref{eq:(C1d)}), $\partial y^n/\partial x^m$, can be obtained directly from (\ref{eq:trans-0_inv})--(\ref{eq:trans-a_inv}):
\begin{eqnarray}
{\partial y^0 \over\partial x^0}
&=& 1 + c^{-2}{\partial \hat{\cal K}\over\partial x^0}  +
c^{-4}{\partial \hat{\cal L}\over\partial x^0}  + {\cal O}(c^{-6}),
 \label{eq:(C1a)_inv}\qquad
{\partial  y^0 \over\partial x^\mu} =
c^{-2} {\partial \hat{\cal K}\over\partial x^\mu}  +
c^{-4}{\partial \hat{\cal L}\over\partial x^\mu}  +
{\cal O}(c^{-5}), \label{eq:(C1b)_inv}\\
{}
{\partial y^\nu  \over \partial x^0} &=&
-\frac{v^\nu_{0}}{c} +
c^{-2} {\partial \hat{\cal Q}^\nu\over\partial x^0} + {\cal O}(c^{-5}), \label{eq:(C1c)_inv}\qquad\qquad\quad\!\!
{\partial y^\nu\over \partial x^\mu}  =
\delta^\nu_\mu  +
c^{-2} {\partial \hat{\cal Q}^\nu\over\partial x^\mu}  + {\cal O}(c^{-4}). \label{eq:(C1d)_inv}
\end{eqnarray}

We note that this Jacobian matrix is composed of the quantities $\partial y^m/\partial x^k$ that are clearly functions of $\{x^k\}$. Using the Jacobian matrix and standard tensor transformation rules, we can express the relationship between the contravariant Minkowski tensor and the contravariant metric of the accelerating frame in the form
{}
\begin{equation}
\eta^{mn}(y)
=\frac{\partial y^m}{\partial x^k}\frac{\partial y^n}{\partial x^l}\gamma^{kl} (x(y)) \qquad {\rm or} \qquad
\eta^{mn}(y(x))=
\frac{\partial y^m}{\partial x^k}\frac{\partial y^n}{\partial x^l}\gamma^{kl} (x).
\label{eq:eta-loc_con**}
\end{equation}
We denote $\hat\eta^{mn}(x)= \eta^{mn}(y(x))$ and, taking the somewhat unusual step of using the contravariant tensor transformation rule (\ref{eq:eta-loc_con**}) together with Eqs.~(\ref{eq:(C1a)_inv})--(\ref{eq:(C1d)_inv}), we obtain explicit expressions for the contravariant components of the accelerated Minkowski metric $\hat\eta^{mn}(x)$, expressed as functions of the global coordinates $\{x^k\}$:
{}
\begin{eqnarray}
\hat\eta^{00}(x)&=&1+\frac{2}{c^2}\Big\{\frac{\partial \hat {\cal K}}{\partial x^0}+ {\textstyle\frac{1}{2}}\gamma^{\epsilon\lambda}\frac{1}{c}
{\partial \hat{\cal K}\over\partial x^\epsilon}\frac{1}{c}
{\partial \hat{\cal K}\over\partial x^\lambda}\Big\}
+\frac{2}{c^4}\Big\{ \frac{\partial \hat {\cal L}}{\partial x^0}+
\gamma^{\epsilon\lambda}\frac{1}{c}
{\partial \hat{\cal K}\over\partial x^\epsilon}
\frac{1}{c}\frac{\partial \hat {\cal L}}{\partial x^\lambda}+
{\textstyle\frac{1}{2}}\Big(\frac{\partial \hat {\cal K}}{\partial x^0}\Big)^2\Big\}+O(c^{-6}), \label{eq:eta00_ctv}\\
{}
\hat\eta^{0\alpha}(x)&=&\frac{1}{c}\Big\{\gamma^{\alpha\epsilon}\frac{1}{c}{\partial \hat{\cal K}\over\partial x^\epsilon}-v_{0}^\alpha\Big\} +
\frac{1}{c^3}\Big\{\gamma^{\alpha\epsilon}\frac{1}{c}\frac{\partial \hat {\cal L}}{\partial x^\epsilon}+
c\frac{\partial \hat {\cal Q}^\alpha}{\partial x^0}+
\gamma^{\epsilon\lambda}\frac{1}{c}{\partial \hat{\cal K}\over\partial x^\epsilon}\frac{\partial \hat {\cal Q}^\alpha}{\partial x^\lambda}-
v^\alpha_{0}\frac{\partial \hat {\cal K}}{\partial x^0}\Big\}+{\cal O}(c^{-5}), \label{eq:eta0a_ctv}\\
{}
\hat\eta^{\alpha\beta}(x)&=&
\gamma^{\alpha\beta}+\frac{1}{c^2}\Big\{
v_{0}^\alpha v_{0}^\beta+\gamma^{\alpha\lambda}
\frac{\partial \hat {\cal Q}^\beta}{\partial x^\lambda}+\gamma^{\beta\lambda}\frac{\partial \hat {\cal Q}^\alpha}{\partial x^\lambda}\Big\}+O(c^{-4}). \label{eq:etaab_ctv}
\end{eqnarray}

As in the case of the direct transformation, we impose the harmonic gauge condition on the metric, to help us establish explicit forms of the transformation functions $\hat{\cal K}$, $\hat{\cal L}$, and $\hat{\cal Q}^\alpha$.

\subsection{Imposing the harmonic gauge condition}

Similarly to the case of direct coordinate transformations, we will use the harmonic gauge conditions. In analogy with the derivation of Eqs.~(\ref{eq:eta_00-cov})--(\ref{eq:eta_ab-cov}), we derive the contravariant metric $\hat\eta^{mn}$:
\begin{eqnarray}
\hat\eta^{00}(x)&=& 1+\frac{2}{c^2}\hat u(x)+\frac{2}{c^4}\hat u^2(x)+O(c^{-6}),
\label{eq:eta_00-ctv}\\
{}
\hat\eta^{0\alpha}(x)&=& \frac{4}{c^3}\hat u^\alpha(x)+O(c^{-5}),
\label{eq:eta_0a-ctv}\\
{}
\hat\eta^{\alpha\beta}(x)&=& \gamma^{\alpha\beta}-\gamma^{\alpha\beta}\frac{2}{c^2}\hat u(x)+O(c^{-4}),
\label{eq:eta_ab-ctv}
\end{eqnarray}
where the inertial scalar $\hat u(x)$ and vector $\hat u^\alpha(x)$ potentials are now expressed via the ``hatted'' transformation functions:
{}
\begin{eqnarray}
\hat u(x)&=& \frac{\partial \hat {\cal K}}{\partial x^0}+
{\textstyle\frac{1}{2}}v_{0}{}_\epsilon v_{0}^\epsilon +\frac{1}{c^2}\Big\{ \frac{\partial \hat {\cal L}}{\partial x^0}+
\frac{v_{0}^\epsilon}{c}\frac{\partial \hat {\cal L}}{\partial x^\epsilon}+
{\textstyle\frac{1}{2}}\Big(\frac{\partial \hat {\cal K}}{\partial x^0}\Big)^2-
\Big(\frac{\partial \hat {\cal K}}{\partial x^0}+
{\textstyle\frac{1}{2}}
v_{0}{}_\epsilon v_{0}^\epsilon\Big)^2\Big\}+{\cal O}(c^{-4}),
\label{eq:pot_loc-w_0-ctv}\\[3pt]
{}
\hat u^\alpha(x)&=& {\textstyle\frac{1}{4}}\Big\{\gamma^{\alpha\epsilon}\frac{1}{c}\frac{\partial \hat {\cal L}}{\partial x^\epsilon}+
c\frac{\partial \hat {\cal Q}^\alpha}{\partial x^0}+v_{0}^\epsilon\frac{\partial \hat {\cal Q}^\alpha}{\partial x^\epsilon}-
v_{0}^\alpha\frac{\partial \hat {\cal K}}{\partial x^0}\Big\}+{\cal O}(c^{-2}).
\label{eq:pot_loc-w_a-ctv}
\end{eqnarray}
One can verify that these two potentials satisfy the following continuity equation in global coordinates $\{x^m\}$:
\begin{equation}
(c\partial_0 +v^\epsilon_0\partial_\epsilon)\hat u+\partial_\epsilon\hat u^\epsilon=
{\cal O}(c^{-2}).  \label{eq:(DeDgGa)cont-eq+}
\end{equation}
In addition, by a direct calculation with the help of Eqs.~(\ref{eq:DeDoGa-K-hat})--(\ref{eq:DeDoGa-Q-hat}) and (\ref{eq:form-inv_0a+})--(\ref{eq:form-inv_ab+}), one can verify that these potentials also satisfy the harmonic equations:
{}
\begin{equation}
\Box_x\hat u={\cal O}(c^{-4}), \qquad
\Delta_x\hat u^\alpha = {\cal O}(c^{-2}),  \label{eq:w-Dalemb-hat}
\end{equation}
where $\Box_x =\gamma^{mn}\partial_m\partial_n$ and $\Delta_x =\gamma^{\epsilon\lambda}\partial_\epsilon\partial_\lambda$ are the d'Alembertian and Laplacian, correspondingly, with respect to $\{x^m\}$. As before, we refer to the metric tensor expressed in terms of $\hat u$ and $\hat u^\alpha$ as the harmonic metric tensor.

Although the expressions for the scalar inertial potentials have different functional dependence on the transformation functions (i.e., (${\cal K},{\cal L},{\cal Q}^\alpha$) vs. ($\hat{\cal K},\hat{\cal L},\hat{\cal Q}^\alpha$)), it is clear that the expressions for $u(y)$ and $\hat u(x)$, given by Eqs.~(\ref{eq:pot_loc-w_0-cov}) and (\ref{eq:pot_loc-w_0-ctv}), represent the same quantity that is being
expressed in terms different coordinates: local $\{y^m\}$ and global $\{x^m\}$, so that $\hat u (x)= u(y(x))$. The same is true for the inertial vector potentials $u^\alpha(y)$ and $\hat u^\alpha(x)$, given by Eqs.~(\ref{eq:pot_loc-w_a-cov}) and (\ref{eq:pot_loc-w_a-ctv}), for which the following relation holds $\hat u^\alpha(x)= u^\alpha(y(x))$.

\subsection{The form of the functions of the harmonic coordinate transformations}

In addition to the constraints provided by Eqs.~(\ref{eq:(DeDgGa)OKs01}) and (\ref{eq:(DeDgGa)OKsa}) on the form of the metric tensor in a moving frame, we can once again derive two additional equations on $\hat{\cal K}$ and $\hat{\cal Q}^\alpha$. From Eqs.~(\ref{eq:(DeDgGa)OKs01}) and (\ref{eq:(DeDgGa)OKsa}) and Eqs.~(\ref{eq:eta00_ctv})--(\ref{eq:etaab_ctv}), we find that these two functions must also satisfy two first order partial differential equations:
{}
\begin{eqnarray}
\gamma^{\alpha\epsilon}\frac{1}{c}{\partial \hat{\cal K}\over\partial x^\epsilon}-v_{0}^\alpha &=&{\cal O}(c^{-4}),
\label{eq:form-inv_0a+} \\
v_{0}^\alpha v_{0}^\beta+\gamma^{\alpha\lambda}
\frac{\partial \hat {\cal Q}^\beta}{\partial x^\lambda}+\gamma^{\beta\lambda}\frac{\partial \hat {\cal Q}^\alpha}{\partial x^\lambda}
+2\gamma_{\alpha\beta}\Big(\frac{\partial \hat {\cal K}}{\partial x^0}+ {\textstyle\frac{1}{2}}\gamma^{\epsilon\lambda}\frac{1}{c}
{\partial \hat{\cal K}\over\partial x^\epsilon}\frac{1}{c}
{\partial \hat{\cal K}\over\partial x^\lambda}\Big) &=& {\cal O}(c^{-2}).
\label{eq:form-inv_ab+}
\end{eqnarray}

In analogy with the derivation of Eq.~(\ref{eq:(DeDg)x0-box}) given in Sec.~\ref{seq:harmform}, we now use the harmonic gauge given by the equation:
\begin{equation}
\Box_{\gamma}y^m= 0,
\label{eq:(DeDg)OK-ctv}
\end{equation}
where $\Box_{\gamma}$ denotes the covariant d'Alembertian with respect to the metric of the inertial frame $\gamma_{mn}(x)$, acting on $y^m(x)$.
Substituting the coordinate transformations (\ref{eq:trans-0_inv})--(\ref{eq:trans-a_inv}) into this equation, we can see that the harmonic gauge conditions restrict the coordinate transformation functions $\hat{\cal K}, \hat{\cal L}$ and $\hat{\cal Q}^\alpha$ only to those that satisfy the following set of second order partial differential equations:
{}
\begin{eqnarray}
\gamma^{\epsilon\lambda}\frac{\partial^2 \hat{\cal K}}{\partial x^\epsilon \partial x^\lambda}&=&{\cal O}(c^{-4}), \label{eq:DeDoGa-K-hat}\\
c^2\frac{\partial^2 \hat{\cal K}}{\partial {x^0}^2}+\gamma^{\epsilon\lambda}\frac{\partial^2 \hat{\cal L}}{\partial x^\epsilon \partial x^\lambda}&=&{\cal O}(c^{-2}), \label{eq:DeDoGa-L-hat}\\
{}
-a_{0}^\alpha+\gamma^{\epsilon\lambda}\frac{\partial^2 \hat{\cal Q}^\alpha}{\partial x^\epsilon \partial x^\lambda}&=&{\cal O}(c^{-2})
\label{eq:DeDoGa-Q-hat}.
\end{eqnarray}
{}
The two sets of partial differential equations on $\hat{\cal K}, \hat{\cal L}$ and $\hat{\cal Q}^\alpha$ given by Eqs.~(\ref{eq:form-inv_0a+})--(\ref{eq:form-inv_ab+}) and (\ref{eq:DeDoGa-K-hat})--(\ref{eq:DeDoGa-Q-hat}) can be used to determine the general structure of these functions. In strict analogy with Eqs.~(\ref{eq:Ka-cov}), (\ref{eq:d-Q}) and (\ref{eq:L-gen-cov}), we can derive the following form for the functions $\hat{\cal K}, \hat{\cal L}$ and $\hat{\cal Q}^\alpha$:
{}
\begin{eqnarray}
\hat{\cal K} (x) &=& \hat\kappa_{0} + c (v^{}_{0\mu}r^\mu)  + {\cal O}(c^{-4}),
\label{eq:K-sum-hat}\\
\hat{\cal L}(x)&=&\hat\ell_{0}+\hat\ell_{0\lambda}\,r^\lambda+
{\textstyle\frac{1}{2}}\hat\ell_{0\lambda\mu}\,r^\lambda r^\mu
+{\textstyle\frac{1}{3}}c
\big(v^{}_{0\epsilon} a_{0}^\epsilon\big)(r^{}_\nu r^\nu)-
{\textstyle\frac{1}{10}}c(\dot{a}_{0\epsilon}r^\epsilon)
(r^{}_\nu r^\nu) - \nonumber\\
&&-
{\textstyle\frac{1}{6}}c^2\frac{\partial}{\partial x^0}\Big(\frac{\partial \hat\kappa_0}{\partial x^0}-
{\textstyle\frac{1}{2}}v_{0}{}_\epsilon v_{0}^\epsilon\Big)(r_\nu r^\nu) +\delta\hat\ell(x)+{\cal O}(c^{-2}),
\label{eq:L-sum-hat}\\
\hat{\cal Q}^\alpha(x)&=& -\hat q^{\alpha}_{0} -
\Big({\textstyle\frac{1}{2}}v^\alpha_{0} v^\epsilon_{0}+
\hat\omega_{0}^{\alpha\epsilon}+\gamma^{\alpha\epsilon}\big(\frac{\partial \hat\kappa_0}{\partial x^0}-
{\textstyle\frac{1}{2}}v^{}_{0\lambda} v_{0}^\lambda\big)\Big){r}_\epsilon -
a^{}_{0\epsilon}\Big(r^\alpha r^\epsilon-
{\textstyle\frac{1}{2}}\gamma^{\alpha\epsilon}{r}_\lambda r^\lambda\Big)+ {\cal O}(c^{-2}),
\label{eq:Q-sum-hat}
\end{eqnarray}
where $r^\alpha$ is defined by
{}
\begin{equation}
r^\alpha=x^\alpha-x^\alpha_0, \qquad x^\alpha_0=z^\alpha_0+c^{-2}{\hat q}^\alpha_0+{\cal O}(c^{-4}),
\label{eq:x_0}
\end{equation}
and $x^\alpha_0(x^0)$ is the vector that connects the origins of the two reference frames, complete to ${\cal O}(c^{-4})$ and expressed as a function of the global time-like coordinate $x^0$, given by Eq.~(\ref{eq:x_0Q}). We can see that $\hat\kappa_0, \hat q^{\alpha}_{0}, \hat\omega_{0}^{\alpha\epsilon}$, $\hat\ell_{0}, \hat\ell_{0\lambda}, \hat\ell_{0\lambda\mu}$ and $\delta\hat\ell(x)$ cannot be determined from the gauge conditions alone. Similarly to the case of direct transformations, we need to apply another set of conditions that would fix the reference frame of a moving observer, which we discuss next.

\subsection{Finding the form of the coordinate transformation functions}

In the case of the metric tensor given by the expressions Eqs.~(\ref{eq:eta_00-ctv})--(\ref{eq:eta_ab-ctv}), the approach discussed in Sec.~\ref{sec:fict-metr} and \ref{sec:fict-metr-eq-mot} that yielded the conditions Eqs.~(\ref{eq:fermi-cov_pot-0})--(\ref{eq:fermi-cov_pot-a}), similarly leads to the following set of equations:
{}
\begin{eqnarray}
 \lim_{|{\vec r}|\rightarrow 0}\hat u(x)     &=&-\hat\varphi_0+
{\cal O}(c^{-4}), \qquad~~~\,\,
\lim_{|{\vec r}|\rightarrow 0} \partial_\beta\hat u(x) = \hat{b}_\beta+  {\cal O}(c^{-4}),
 \label{eq:fermi-ctv_0}\\
 \lim_{|{\vec r}|\rightarrow 0}\hat u^\alpha(x)     &=&
{\cal O}(c^{-2}),  \qquad\qquad\qquad
\lim_{|{\vec r}|\rightarrow 0} \partial_\beta\hat u^\alpha(x)     =
{\textstyle\frac{1}{3}}\delta^\alpha_\beta\dot{\hat\varphi}_0+
{\cal O}(c^{-2}),
 \label{eq:fermi-ctv}
\end{eqnarray}
where, similarly to Eq.~(\ref{eq:inert-accel}), $\hat\varphi_0(x^0)=\varphi_0\big(x^0+c^{-2}\hat\kappa_0+{\cal O}(c^{-4})\big)$ is the background potential at the observer's world-line (where we used $y^0(x^0)=x^0+c^{-2}\hat\kappa_0(x^0)+{\cal O}(c^{-4})$, valid on the world-line) and the measured acceleration $\hat{b}^\alpha$ is related to the coordinate acceleration $\ddot x^\alpha_{0}$, given by Eq.~(\ref{eq:geod_eq-local-cov}), as $\hat{b}^\alpha(x^0)=\big(1-2\varphi_0/c^2+{\cal O}(c^{-4})\big)\ddot x^\alpha_{0}\big(y^0(x^0)\big)$, so that $\hat{b}^\alpha$ relates to ${b}^\alpha$ as
{}
\begin{eqnarray}
\hat{b}^\alpha(x^0)&=& b_\epsilon\big(x^0+c^{-2}\hat\kappa_0+{\cal O}(c^{-4})\big)\Big\{\gamma^{\alpha\epsilon}-\frac{1}{c^2}\Big(
{\textstyle\frac{1}{2}}v^\alpha_{0}v_0^\epsilon +\omega_{0}^{\alpha\epsilon}-\gamma^{\alpha\epsilon}\varphi_0\Big)\Big\}
+\frac{1}{c^2} v^\alpha_0\dot\varphi_0+ {\cal O}(c^{-4}).
\label{eq:geod_eq-local-cov+=}
\end{eqnarray}

Imposing the conditions Eqs.~(\ref{eq:fermi-ctv_0})--(\ref{eq:fermi-ctv}) on the potentials $\hat u$ and $\hat u^\alpha$ given by Eqs.~(\ref{eq:eta_00-ctv})--(\ref{eq:eta_ab-ctv}) results in a set of partial differential equations set on the world-line of the accelerated observer. These equations can used to determine the coordinate transformation functions entering Eqs.~(\ref{eq:K-sum-hat})--(\ref{eq:Q-sum-hat}) which were found to be:
{}
\begin{eqnarray}
\hat\kappa_0 &=&-\int_{x^0_0}^{x^0}\!\!\!
\Big(\hat\varphi_0-{\textstyle\frac{1}{2}}v^{}_{0\epsilon} v_{0}^{\epsilon}-c^{-2}v^{}_{0\epsilon} \dot{\hat q}_{0}^\epsilon\Big)dx'^0 + {\cal O}(c^{-4}), \label{eq:dir0-K-ctv}\\
a_{0}^\alpha&=&b^{[0]\alpha} +{\cal O}(c^{-4}).
\label{eq:a-Newton-ctv}
\end{eqnarray}
In addition, the use of Eqs.~(\ref{eq:fermi-ctv_0})--(\ref{eq:fermi-ctv}) yields the following solutions for the functions ${\hat\ell}_{0}$, $\dot{\hat\ell}^\alpha_0$, $\hat\ell^\alpha_{0}$, and $\hat\ell^{\alpha\lambda}_{0}$:
{}
\begin{eqnarray}
{\textstyle\frac{1}{c}}\dot{\hat\ell}_{0}&=&-
{\textstyle\frac{1}{8}}(v_{0\epsilon}v^\epsilon_{0})^2-
{\textstyle\frac{1}{2}}(v_{0\epsilon}v^\epsilon_{0})\hat\varphi_0+
{\textstyle\frac{1}{2}} \hat\varphi_0^2+{\cal O}(c^{-2}),\\[2pt]
\label{eq:ell_0-hat}
{\textstyle\frac{1}{c}}\dot{\hat\ell}^\alpha_{0}&=&\hat{b}^{[2]\alpha}+\big({\textstyle\frac{1}{2}}(v_{0\epsilon}v^\epsilon_{0})-\hat\varphi_0\big)a^\alpha_0+
{\cal O}(c^{-2}),\\[2pt]
\label{eq:ell-dot-hat}
{\textstyle\frac{1}{c}}\hat\ell^\alpha_{0}&=&{\dot {\hat q}}^\alpha_{0}-
{\textstyle\frac{1}{2}}v_0^\alpha\big(v^{}_{0\epsilon}v^\epsilon_{0}\big)-v^\alpha_0\hat\varphi_0+{\cal O}(c^{-2}),\\[2pt]
\label{eq:ell-hat}
{\textstyle\frac{1}{c}}\hat\ell^{\alpha\beta}_{0}&=&
v^\alpha_{0}a^\beta_{0}+v^\beta_{0}a^\alpha_{0}-{\textstyle\frac{2}{3}}\gamma^{\alpha\beta}v_{0\epsilon}a_{0}^\epsilon+{\cal O}(c^{-2}).
\label{eq:ellb-dd-hat}
\end{eqnarray}
Using the same argument concerning the symmetry properties of $\hat\ell^{\alpha\beta}_0$ that led to Eq.~(\ref{eq:omega}), we find the following unique choice for the anti-symmetric matrix  ${\dot{\hat \omega}}_{0}^{\alpha\beta}$ representing the Thomas precession:
\begin{equation}
\dot{\hat \omega}_{0}^{\alpha\beta}=-\dot\omega_{0}^{\alpha\beta}=
-{\textstyle\frac{1}{2}}\big(v^{\alpha}_{0}a^{\beta}_{0}-v^{\beta}_{0}a^{\alpha}_{0}\big)+{\cal O}(c^{-2}).
\label{eq:omega-hat}
\end{equation}

Finally, Eqs.~(\ref{eq:ell-hat}) and (\ref{eq:ell-dot-2}) allow us to determine the equation for $\hat q_{0}^{\alpha}$. Indeed, differentiating  Eq.~(\ref{eq:ell-hat}) with respect to time and subtracting the result from Eq.~(\ref{eq:ell-dot-hat}), we obtain:
{}
\begin{eqnarray}
\ddot {\hat q}_{0}^{\alpha}&=&  \hat{b}^{[2]\alpha} +\big(v^\alpha_{0}a_0^\epsilon+a^\alpha_{0}v_0^\epsilon\big)v^{}_{0\epsilon}+v^\alpha_{0}\dot{\hat\varphi}_0+
{\cal O}(c^{-2}).
\label{eq:q-ddot-hat}
\end{eqnarray}

By combining (\ref{eq:a-Newton-ctv}) and (\ref{eq:q-ddot-hat}), we obtain the equations of motion of the accelerated observer with respect to $\{x^m\}$:
\begin{eqnarray}
\ddot x^\alpha_0(x^0)=  a^\alpha_{0}+c^{-2}\ddot {\hat q}_{0}^{\alpha} +
{\cal O}(c^{-4})&=&
b^{[0]\alpha}+\frac{1}{c^2}\Big\{\hat{b}^{[2]\alpha} +\big(v^\alpha_{0}a_0^\epsilon+a^\alpha_{0}v_0^\epsilon\big)v^{}_{0\epsilon}+v^\alpha_{0}\dot{\hat\varphi}_0\Big\}+
{\cal O}(c^{-4})\nonumber\\
&=& \hat{b}_\epsilon\Big\{\gamma^{\alpha\epsilon}+\frac{1}{c^2}\big(v^\alpha_{0}v_0^\epsilon+\gamma^{\alpha\epsilon}v^{}_{0\mu}v_0^\mu \big)\Big\}+ \frac{1}{c^2}v^\alpha_{0}\dot{\hat\varphi}_0+
{\cal O}(c^{-4}).
\label{eq:geod_eq-local-ctv}
\end{eqnarray}

The equation of motion (\ref{eq:geod_eq-local-ctv}) establish the correspondence between the external force acting on the test particle, $\hat{b}^\alpha$, and the resulting acceleration of the moving frame $\ddot x^\alpha_0(x^0)$ as measured by the inertial observer.

\subsection{Summary of results for the inverse transformation}

The coordinate transformations between the global coordinates of the inertial frame and the local coordinates introduced in the proper reference frame of an accelerating observer have the following form
{}
\begin{eqnarray}
y^0&=& x^0+c^{-2}\hat{\cal K}(x^0,x^\epsilon)+c^{-4}\hat{\cal L}(x^0,x^\epsilon)+{\cal O}(c^{-6}),\\[3pt]
y^\alpha&=& x^\alpha-z^\alpha_{0}(x^0)+c^{-2}\hat{\cal Q}^\alpha(x^0,x^\epsilon)+{\cal O}(c^{-4}),
\end{eqnarray}
with the transformation functions $\hat{\cal K}, \hat{\cal L}$ and $\hat{\cal Q}^\alpha$ given by
{}
\begin{eqnarray}
\hat{\cal K}(x)&=& -\int_{x^0_0}^{x^0}\!\!\!
\Big(\hat\varphi_0-{\textstyle\frac{1}{2}}v^{}_{0\epsilon} v_{0}^{\epsilon}-c^{-2}v^{}_{0\epsilon} \dot{\hat q}_{0}^\epsilon\Big)dx'^0+
c (v^{}_{0\epsilon} r^\epsilon) + {\cal O}(c^{-4}),\\
{}
\hat{\cal L}(x)&=& \!\int_{x^0_{0}}^{x^0}\!\!\!
 \Big(\!-{\textstyle\frac{1}{8}}
\big(v^{}_{0\epsilon} v_{0}^\epsilon\big)^2-
{\textstyle\frac{1}{2}}(v^{}_{0\epsilon} v_{0}^\epsilon)\hat\varphi_0+
{\textstyle\frac{1}{2}}\hat\varphi_0^2\Big)dx'^0+
c\big({\dot {\hat q}}^{}_{0\lambda}-
{\textstyle\frac{1}{2}} v^{}_{0\lambda}(v^{}_{0\epsilon}v^\epsilon_{0})-v_{0\lambda}\hat\varphi_0\big)r^\lambda+\nonumber\\
&&\hskip 20pt+~
{\textstyle\frac{1}{2}}c
(v^{}_{0\epsilon}a^{}_{0\lambda} + v^{}_{0\lambda}a^{}_{0\epsilon}
+{\textstyle\frac{1}{3}}\gamma_{\epsilon\lambda}\dot{\hat\varphi}_0)r^\epsilon r^\lambda -
{\textstyle\frac{1}{10}}c(\dot{a}_{0\epsilon}r^\epsilon)(r_\nu r^\nu)+\delta\hat\ell(x)+{\cal O}(c^{-2}), \label{eq:suma_hatL}\\[3pt]
\hat{\cal Q}^\alpha(x)&=& -{\hat q}^{\alpha}_{0} -
\Big({\textstyle\frac{1}{2}}v^\alpha_{0} v^\epsilon_{0}+
\hat\omega_{0}^{\alpha\epsilon}-\gamma^{\alpha\epsilon}\hat\varphi_0\Big){r}_\epsilon -
{a_{0}}_\epsilon\Big(r^\alpha r^\epsilon-{\textstyle\frac{1}{2}}
\gamma^{\alpha\epsilon}{r}_\lambda r^\lambda\Big)
+ {\cal O}(c^{-2}),
\end{eqnarray}
where $r^\alpha=x^\alpha-x^\alpha_0$ is given by Eq.~(\ref{eq:x_0}), the anti-symmetric relativistic precession matrix $\hat\omega_{0}^{\alpha\lambda}$ is given by Eq.~(\ref{eq:omega-hat}), and the post-Newtonian component of the spatial coordinate origin of the local frame, $\hat q_{0}^{\alpha}$ is given by Eq.~(\ref{eq:q-ddot-hat}).

Substituting these solutions for the functions $\hat{\cal K}$, $\hat{\cal L}$ and $\hat{\cal Q}^\alpha$ into the expressions for the inertial potentials $u$ and $u^\alpha$ given by Eqs.~(\ref{eq:pot_loc-w_0-ctv})--(\ref{eq:pot_loc-w_a-ctv}), we find the following form for these potentials:
{}
\begin{eqnarray}
\hat u(x)&=&({\hat a}_\epsilon r^\epsilon)-\hat\varphi_0+\frac{1}{c^2}\Big\{
{\textstyle\frac{1}{2}}\Big({\dot a}^{}_{0\epsilon} v^{}_{0\lambda}
+v^{}_{0\epsilon}{\dot a}^{}_{0\lambda}  +a^{}_{0\epsilon} a^{}_{0\lambda}+{\textstyle\frac{1}{3}}\gamma_{\epsilon\lambda}\ddot{\hat\varphi}_0\Big)r^\epsilon r^\lambda -
{\textstyle\frac{1}{10}}
(\ddot a^{}_{0\epsilon} r^\epsilon)(r_\mu r^\mu)\Big\}+\nonumber\\
&&\hskip 200pt+
(c\partial_0+ v^\epsilon_0\partial_\epsilon) {\textstyle\frac{1}{c}}\delta\hat\ell+{\cal O}(c^{-4}),
\label{eq:pot_loc-w_0-ctv+}\\[3pt]
\hat u^\alpha(x)&=&-
{\textstyle\frac{1}{10}}\big\{3r^\alpha r^\epsilon-
\gamma^{\alpha\epsilon}r_\mu r^\mu\big\}{\dot a}^{}_{0\epsilon}+
{\textstyle\frac{1}{3}}\dot{\hat\varphi}_0 r^\alpha+\partial^\alpha {\textstyle\frac{1}{4c}}\delta\hat\ell+{\cal O}(c^{-2}),
\label{eq:pot_loc-w_a-ctv+}
\end{eqnarray}
where ${\hat a}^\alpha$ denotes the frame-reaction acceleration, which is equal to the  external acceleration ${\hat b}^\alpha$ given by Eq.~(\ref{eq:geod_eq-local-cov+=}) and measured in the global frame; i.e., ${\hat a}^\alpha(x^0)\equiv {\hat b}^\alpha(x^0)=b^{[0]\alpha}+c^{-2}{\hat b}^{[2]\alpha}+{\cal O}(c^{-4})$.

The same argument that allowed us to eliminate $\delta\ell$ in Sec.~\ref{sec:sumdirect} works here, allowing us to omit $\delta\hat\ell$. Therefore, we can now present the metric of the moving observer expressed in the global coordinates $\{x^n\}$. Substituting the expressions for the inertial potentials given by (\ref{eq:pot_loc-w_0-ctv+}) and (\ref{eq:pot_loc-w_a-ctv+}) into Eqs.~(\ref{eq:eta_00-ctv})--(\ref{eq:eta_ab-ctv}), and omitting $\delta\ell$, we have:
{}
\begin{eqnarray}
\hat \eta^{00}(x)&=& 1+\frac{2}{c^2}\Big\{({\hat a}_\epsilon r^\epsilon)-\hat\varphi_0\Big\}+\frac{2}{c^4}\Big\{\big((a^{}_{0\epsilon} r^\epsilon)-\hat\varphi_0\big)^2+\nonumber\\*
&&\hskip 65pt+~{\textstyle\frac{1}{2}}\Big({\dot a}^{}_{0\epsilon} v^{}_{0\lambda}
+v^{}_{0\epsilon}{\dot a}^{}_{0\lambda}  +a^{}_{0\epsilon} a^{}_{0\lambda}+{\textstyle\frac{1}{3}}\gamma_{\epsilon\lambda}\ddot{\hat\varphi}_0\Big)r^\epsilon r^\lambda -
{\textstyle\frac{1}{10}}
(\ddot a^{}_{0\epsilon} r^\epsilon)(r_\mu r^\mu)\Big\}+{\cal O}(c^{-6}),
\label{eq:eta_00-ctv-fin}\\
\hat \eta^{0\alpha}(x)&=& -\frac{4}{c^3}
\Big\{{\textstyle\frac{1}{10}}\big(3r^\alpha r^\epsilon-
\gamma^{\alpha\epsilon}r_\mu r^\mu\big){\dot a}^{}_{0\epsilon}-\textstyle{\frac{1}{3}}{\dot{\hat\varphi}}_0 r^\alpha\Big\}
+{\cal O}(c^{-5}),
\label{eq:eta_0a-ctv-fin}\\
\hat \eta^{\alpha\beta}(x)&=& \gamma^{\alpha\beta}-\gamma^{\alpha\beta}\frac{2}{c^2} \Big\{(a^{}_{0\epsilon} r^\epsilon)-\hat\varphi_0\Big\}+{\cal O}(c^{-4}).
\label{eq:eta_ab-ctv-fin}
\end{eqnarray}
The coordinate transformations that put the observer in this reference frame are given by:
{}
\begin{eqnarray}
y^0&=& x^0+c^{-2}\Big\{-\!\!\int_{x^0_{0}}^{x^0}\!\!\!
\Big(\hat\varphi_0-{\textstyle\frac{1}{2}}(v^{}_{0\epsilon} +c^{-2}\dot{\hat q}_{0\epsilon})(v_{0}^\epsilon+c^{-2}\dot{\hat q}_{0}^\epsilon)+c^{-2}\big\{{\textstyle\frac{1}{8}}(v^{}_{0\epsilon}v^\epsilon_{0})^2+
{\textstyle\frac{1}{2}}(v^{}_{0\epsilon} v_{0}^\epsilon)\hat\varphi_0-
{\textstyle\frac{1}{2}}\hat\varphi_0^2\big\}\Big)dx'^0 +\nonumber\\
&&\hskip 50pt{}
+c\,\Big(1-c^{-2}\big({\textstyle\frac{1}{2}} (v^{}_{0\mu}v^\mu_{0}) +\hat\varphi_0\big)\Big)\Big(v^{}_{0\epsilon} +c^{-2}{\dot {\hat q}}^{}_{0\epsilon}\Big)r^\epsilon\Big\}
+\nonumber\\
&&~~~{}+c^{-4}\Big\{{\textstyle\frac{1}{2}}c\big(v^{}_{0\epsilon}a^{}_{0\lambda}+a^{}_{0\epsilon}v^{}_{0\lambda}+{\textstyle\frac{1}{3}}\gamma_{\epsilon\lambda}\dot{\hat\varphi}_0\big)r^\epsilon r^\lambda-{\textstyle\frac{1}{10}}c(\dot{a}_{0\epsilon}r^\epsilon)(r_\nu r^\nu)\Big\}+{\cal O}(c^{-6}),
\label{eq:transfrom-fin-0-ctv}\\
y^\alpha&=& r^\alpha-
c^{-2}\Big\{\Big({\textstyle\frac{1}{2}}v^\alpha_{0} v^\epsilon_{0}+
\hat\omega_{0}^{\alpha\epsilon}-\gamma^{\alpha\epsilon}\hat\varphi_0\Big)r_\epsilon +
{a_{0}}_\epsilon\Big(r^\alpha r^\epsilon-
{\textstyle\frac{1}{2}}\gamma^{\alpha\epsilon}{r}_\lambda r^\lambda\Big)\Big\}+{\cal O}(c^{-4}),
\label{eq:transfrom-fin-a-ctv}
\end{eqnarray}
where $r^\alpha=x^\alpha-x^\alpha_0(t)$, as defined by Eq.~(\ref{eq:x_0}).
Eq.~(\ref{eq:Q-hat}) suggests that
$
\hat q^\alpha_0=q^\alpha_0-(v^\alpha_0/c)\int (\varphi_0-
{\textstyle\frac{1}{2}}v^{}_{0\epsilon} v_{0}^{\epsilon})dy'^0  + {\cal O}(c^{-2}),
$
which, after accounting for Eq.~(\ref{eq:q-dot}), in the case of constant velocity motion and the absence of the background external potential $\hat\varphi_0=0$, results in ${\dot {\hat q}}_0^\alpha={\cal O}(c^{-2})$. Consequently, in this case the metric given by Eqs.~(\ref{eq:eta_00-ctv-fin})--(\ref{eq:eta_ab-ctv-fin}) reduces to the Minkowski metric, $\eta^{mn}=\gamma^{mn}$, and the transformations above reduce to the Lorentz transformations between the coordinates of the accelerated $(y^0,y^\alpha)$ and inertial $(x^0,x^\alpha)$ reference frames:
{}
\begin{eqnarray}
y^0&=& \Big(1+c^{-2}{\textstyle\frac{1}{2}}(v^{}_{0\epsilon} v_{0}^\epsilon)-c^{-4}{\textstyle\frac{1}{8}}(v^{}_{0\epsilon}v^\epsilon_{0})^2\Big)x^0+
\Big(1 - c^{-2}{\textstyle\frac{1}{2}}
v^{}_{0\epsilon}v^\epsilon_{0}\Big)c^{-1}(v^{\lambda}_{0}r^\epsilon)
+{\cal O}(c^{-6}),
\label{eq:transfrom-fin-0-ctv-red}\\
y^\alpha&=& \Big(\delta^\alpha_\epsilon-
c^{-2}{\textstyle\frac{1}{2}}v^\alpha_{0} v^{}_{0\epsilon}\Big)r^\epsilon +{\cal O}(c^{-4}).
\label{eq:transfrom-fin-a-ctv-red}
\end{eqnarray}
If we now write $x^\alpha_0(t)=v^\alpha_0 c^{-1}x^0$, we obtain expressions identical to Eqs.~(\ref{eq:1.12a})--(\ref{eq:1.12b}), demonstrating correspondence of our results to those established previously using different techniques.

This set of results concludes our derivation of the coordinate transformations from local coordinates of an accelerated reference frame to the global coordinates introduced in the inertial frame.

\section{Discussion and Conclusions}
\label{sec:end}

In this paper we introduced a new approach to construct coordinate transformations between inertial and accelerating reference frames. Our objective was to establish properties of a frame associated with a given accelerating world-line and to find the metric tensor corresponding to this frame; we achieved this goal by imposing a set of clearly defined coordinate and physical conditions. Specifically, we combined the use of the harmonic gauge and physical insight based on the dynamical properties of the proper reference frame of an accelerating observer. This way, we were able to find a unique non-rotating, accelerating coordinate system that corresponds to a given accelerating world-line. We derived an explicit form of the coordinate transformation between the inertial and accelerating reference frames to the first post-Galilean order. The transformations remain valid so long as $v\ll c$. In addition to the direct transformation, which allows us to express the coordinates $\{x^m\}$ of the inertial reference frame in terms of the accelerating coordinates $\{y^m\}$, we also developed the inverse coordinate transformations in explicit form.

The metric associated with the proper reference frame of an accelerated observer has been studied by many researchers (see for instance, \cite{Synge-1960,MTW,Ni-1977,Ni-Zimmermann-1978}). It was found that when only the acceleration $a_0^\alpha$ of the reference frame is taken into account, the corresponding metric in local non-rotating coordinates $\{y'^m\}\equiv(y'^0,y'^\alpha)$ takes the form \cite{Moeller-book-1972,MTW}:
\begin{equation}
\eta^S_{00}(y')=\big(1+\frac{1}{c^2}(a^{}_{0\epsilon} y'^\epsilon)\big)^2, \qquad \eta^S_{0\alpha}(y')=0, \qquad \eta^S_{\alpha\beta}(y')=\gamma_{\alpha\beta}.
\label{eq:Synge}
\end{equation}
To establish the correspondence of the newly found metric $\eta_{mn}(y)$ of an accelerated observer in harmonic coordinates of its proper reference frame Eqs.~(\ref{eq:eta_00-cov-fin+})--(\ref{eq:eta_ab-cov-fin+}) to the metric $\eta^S_{mn}(y')$ given in the form of Eq.~(\ref{eq:Synge}) we can use the same approach employing the functional ${\cal KLQ}$-parameterization. To do this, first of all, we note that in the absence of the external potential ($\varphi_0=0$) the metric Eqs.~(\ref{eq:eta_00-cov-fin+})--(\ref{eq:eta_ab-cov-fin+}) reduces to:
{}
\begin{eqnarray}
\eta_{00}(y)&=& 1-\frac{2}{c^2}(a_\epsilon y^\epsilon)+\frac{2}{c^4}\Big\{
{\textstyle\frac{1}{2}}\Big(\gamma_{\epsilon\lambda} a^{}_{0\mu} a^{\mu}_{0}-a^{}_{0\epsilon}a^{}_{0\lambda}\Big)
y^\epsilon y^\lambda+
{\textstyle\frac{1}{10}}
(\ddot a^{}_{0\epsilon} y^\epsilon)(y_\mu y^\mu)\Big\}+{\cal O}(c^{-6}),
\label{eq:eta_00-cov-fin+++}\\
\eta_{0\alpha}(y)&=& \gamma_{\alpha\lambda}\frac{4}{c^3}\Big\{
{\textstyle\frac{1}{10}}\big(3y^\lambda y^\epsilon-
\gamma^{\lambda\epsilon}y_\mu y^\mu\big){\dot a}_{0\epsilon}\Big\}+{\cal O}(c^{-5}),
\label{eq:eta_0a-cov-fin+++}\\
\eta_{\alpha\beta}(y)&=& \gamma_{\alpha\beta}+\gamma_{\alpha\beta}\frac{2}{c^2} (a^{}_{0\epsilon} y^\epsilon)+{\cal O}(c^{-4}).
\label{eq:eta_ab-cov-fin+++}
\end{eqnarray}
Then, in analogy to Eqs.~(\ref{eq:trans-0})--(\ref{eq:trans-a}) we can introduce coordinate transformations $y^m=f^m(y'^k)$, which would depend on a set of to-be-determined functions of ${\cal K}', {\cal L}',$ and ${\cal Q}'^\alpha$. Matching the two metric tensors via the usual tensor transformation rule $\eta_{mn}(y)=({\partial y'^k}/{\partial y^m})({\partial y'^l}/{\partial y^n})\eta^S_{kl}(y')$ yields unique solutions for these functions. As a result, we can obtain the following coordinate transformation that transforms the well-known metric of an accelerated observer in coordinates (\ref{eq:Synge}) to the new harmonic accelerated metric given by Eqs.~(\ref{eq:eta_00-cov-fin+++})--(\ref{eq:eta_ab-cov-fin+++}):
\begin{equation}
y'^0=y^0+c^{-4} \big\{c{\textstyle\frac{1}{10}} (\dot{a}_{0\epsilon} y^\epsilon)(y^{}_\lambda y^\lambda)\big\},~~~~
y'^\alpha=y^\alpha + c^{-2} \big\{a^{}_{0\epsilon} (y^\alpha y^\epsilon -
{\textstyle\frac{1}{2}} \gamma^{\alpha\epsilon} y^{}_\lambda y^\lambda) \big\}.
\end{equation}
Note that the inverse transformations are obtained simply by replacing $(y'^0, y'^\alpha)$ with $(y^0, y^\alpha)$ and the acceleration $a^\alpha_0$ with $-a^\alpha_0$. This establishes correspondence between our results reported in \cite{Synge-1960,MTW,Ni-1977,Ni-Zimmermann-1978} that were previously derived under different gauge conditions and using different techniques for constricting the proper reference frame of an accelerated observer. The resulting metric (\ref{eq:eta_00-cov-fin+++})--(\ref{eq:eta_ab-cov-fin+++}) can be used to study physical phenomena in this frame.

The new results reported in this paper find a good correspondence with the results previously obtained by other researchers who also used the harmonic gauge conditions. Thus, the coordinate transformations that we derived in the context of a flat space-time find their exact correspondence with the results established in the presence of gravity for both direct and inverse coordinate transformations given in  \cite{DSX-I,DSX-II,DSX-III,DSX-IV} and  \cite{Brumberg-Kopejkin-1989,Kopejkin-1988,Brumberg-Kopejkin-1989-2,Kopeikin:2004ia}  correspondingly. However, our approach allows one to consistently and within the same framework develop both direct and inverse transformations, corresponding equations of motion, and explicit forms of the metric tensors in the various reference frames involved. The difficulty of this task was mentioned in \cite{Soffel:2003cr} when the post-Newtonian motion of a gravitational $N$-body system was considered; our proposed formulation successfully resolves this important issue. As an added benefit, the new approach provides one with a good justification to eliminate the functions $\delta \kappa$, $\delta \xi$ and $\delta \ell$ yielding a complete form for the transformation functions ${\cal K}$, ${\cal L}$ and ${\cal Q}^\alpha$ involved  in the  transformations (as well as their ``hatted'' counterparts).

The significance of our result is that for the first time, a formalism for the coordinate transformation between inertial and accelerating reference frames is provided, presenting both the direct and inverse transformations in explicit form.
By combining inverse and direct transformations, the transformation rules between arbitrary accelerating frames can be obtained. Furthermore, it is possible to combine direct (or inverse) transformations, and obtain another transformation that can be represented by our formalism, as shown explicitly in \cite{Turyshev-96}. This leads to an approximate finite group structure that extends the Poincar\'e group of transformations to accelerating reference frames.

The results obtained in this paper provide a clear framework to describe observables, field transformations, and corresponding equations of motion that are needed to describe modern-day high-precision experiments. Specifically, the new approach is designed to facilitate the analysis of relativistic phenomena where effects of acceleration may be significant. We should note that the approach we presented can be used in an iterative manner: if greater accuracy is desired, the coordinate transformations (\ref{eq:trans-0})--(\ref{eq:trans-a}) can be expanded to include higher-order terms. Furthermore, the same approach relying on the functional ${\cal KLQ}$-parameterization may be successfully applied to the case of describing the gravitational dynamics of an astronomical $N$-body system and dynamically-rotating reference frames. This  work has begun and the results, when available, will be reported elsewhere.

\begin{acknowledgments}
We thank Sami W. Asmar, Curt J. Cutler, William M. Folkner,
Michael M. Watkins, and James G. Williams of JPL for their interest and support during the work and preparation of this manuscript. We also thank Sergei M. Kopeikin for his insightful comments and suggestions. This work was performed at the Jet Propulsion Laboratory, California Institute of Technology, under a contract with the National Aeronautics and Space Administration.

\end{acknowledgments}

\end{document}